%% file: ppisn_CSM.tex
\documentclass{aa}              
\usepackage{graphicx}
\usepackage[varg]{txfonts}
\usepackage[
colorlinks,citecolor=blue,linkcolor=blue,urlcolor=blue]{hyperref}
\usepackage{amsmath}
\usepackage{morefloats}
\usepackage[dvipsnames]{xcolor}
\usepackage{footnote}
\usepackage{multirow}
\usepackage{chngpage}
\usepackage{lscape}
\usepackage{url}
\usepackage{float}
\usepackage[scientific-notation=true]{siunitx}
\usepackage{subfigure}
\usepackage{diagbox} 
\usepackage{supertabular}

\newcommand{\udef}{\stackrel{\mathrm{def}}{=}}
\newcommand{\pair}{\ensuremath{e^\pm}}

\newcommand{\kms}{{\mathrm{km\ s^{-1}}}}

\DeclareRobustCommand{\Eqref}[1]{Equation~\ref{#1}}
\DeclareRobustCommand{\Figref}[1]{Figure~\ref{#1}}
\DeclareRobustCommand{\Tabref}[1]{Table~\ref{#1}}
\DeclareRobustCommand{\Secref}[1]{Section~\ref{#1}}

\DeclareRobustCommand{\eqref}[1]{\Eqref{#1}}
\DeclareRobustCommand{\figref}[1]{\Figref{#1}}

\DeclareRobustCommand{\secref}[1]{\Secref{#1}}

\begin{document}


\title{Predictions for the hydrogen-free ejecta of pulsational pair-instability supernovae}

\titlerunning{CSM from Pulsational Pair-Instability}
  
\author{M.~Renzo \inst{1,2} 
        \and
        R.~Farmer \inst{1,6}
        \and
        S.~Justham \inst{3,4,1}
        \and
        Y.~G\"otberg \inst{5}
        \and
        S~.E.~de~Mink \inst{6,1}
        \and
        E.~Zapartas \inst{7}
        \and
        P.~Marchant \inst{8,9}
        \and
        N.~Smith \inst{10}
}

\institute{
  {$^1$  Anton Pannekoek Institute for Astronomy and GRAPPA, University of
    Amsterdam, NL-1090 GE Amsterdam, The Netherlands}\\
  {$^2$  Center for Computational Astrophysics, Flatiron Institute,
    New York, NY 10010, USA}\\
  {$^3$  School of Astronomy \& Space Science, University of the Chinese Academy of Sciences, Beijing 100012, China}\\
  {$^4$  National Astronomical Observatories, Chinese Academy of Sciences, Beijing 100012, China}\\
  {$^5$  The observatories of the Carnegie institution for science, 813 Santa Barbara St., Pasadena, CA 91101, USA}\\
  {$^6$  Center for Astrophysics, Harvard \& Smithsonian, 60 Garden Street, Cambridge, MA 02138, USA}\\
  {$^7$  Geneva Observatory, University of Geneva, CH-1290 Sauverny, Switzerland}\\
  {$^8$  Department of Physics and Astronomy, Northwestern University, 2145 Sheridan Road, Evanston, IL 60208, USA}\\
  {$^9$  Institute of Astrophysics, KU Leuven,Celestijnlaan 200D, 3001 Leuven, Belgium}\\
  {$^{10}$ Steward Observatory, University of Arizona, 933 N. Cherry Ave., Tucson, AZ 85721, USA}\\
\email{\href{mailto:mrenzo@flatironinstitute.org}{mrenzo@flatironinstitute.org}}
}

\abstract{Present and upcoming time-domain astronomy efforts, in part
  driven by gravitational-wave follow-up campaigns, will unveil a
  variety of rare explosive transients in the sky. Here, we focus on
  pulsational pair-instability evolution, which can result in
  signatures that are observable with electromagnetic and gravitational
  waves. We simulated grids of bare helium stars to characterize
  the resulting black hole (BH) masses together with the ejecta composition,
  velocity, and thermal state. We find that the stars do not react
  ``elastically'' to the thermonuclear ignition in the core: there is
  not a one-to-one correspondence between pair-instability driven
  ignition and mass ejections, which causes ambiguity as to what is an
  observable pulse. In agreement with previous studies, we find that
  for initial helium core masses of
  $37.5\,M_\odot\lesssim M_\mathrm{He,init}\lesssim41\,M_\odot$,
  corresponding to carbon-oxygen core masses
  $27.5\,M_\odot\lesssim M_\mathrm{CO}\lesssim30.1\,M_\odot$, the
  explosions are not strong enough to affect the surface.  With
  increasing initial helium core mass, they become progressively
  stronger causing first large radial expansion
  ($41\,M_\odot \lesssim M_\mathrm{He,init}\lesssim42\,M_\odot$,
  corresponding to
  $30.1\,M_\odot\lesssim M_\mathrm{CO}\lesssim30.8\,M_\odot$) and,
  finally, also mass ejection episodes (for
  $M_\mathrm{He,init}\gtrsim42\,M_\odot$, or
  $M_\mathrm{CO}\gtrsim30.8\,M_\odot$). The lowest mass helium core to
  be fully disrupted in a pair-instability supernova is
  $M_\mathrm{He,init}\simeq80\,M_\odot$, corresponding to
  $M_\mathrm{CO}\simeq55\,M_\odot$. Models with
  $M_\mathrm{He,init}\gtrsim200\,M_\odot$
  ($M_\mathrm{CO}\gtrsim 114\,M_\odot$) reach the photodisintegration
  regime, resulting in BHs with masses of
  $M_\mathrm{BH}\gtrsim125\,M_\odot$. Although this is currently
    considered unlikely, if BHs from these models form via (weak) explosions, the previously-ejected material might be
  hit by the blast wave and convert kinetic energy into observable
  electromagnetic radiation. We characterize the hydrogen-free
  circumstellar material from the pulsational pair-instability of
  helium cores by simply assuming that the ejecta maintain a constant
  velocity after ejection. We find that our models produce helium-rich
  ejecta with mass of
  $10^{-3}\,M_\odot\lesssim M_\mathrm{CSM}\lesssim40\,M_\odot$, the
  larger values corresponding to the more massive progenitor
  stars. These ejecta are typically launched at a few thousand $\kms$
  and reach distances of $\sim$\,$10^{12}-10^{15}\,\mathrm{cm}$ before
  the core-collapse of the star. The delays between mass ejection
  events and the final collapse span a wide and mass-dependent range
  (from subhour to $10^{4}$ years), and the shells ejected can also
  collide with each other, powering supernova impostor events before
  the final core-collapse. The range of properties we find suggests a
  possible connection with (some) type Ibn supernovae. }

\maketitle
\section{Introduction}
\label{sec:intro}

\begin{figure*}[hbtp]
  \centering
  \includegraphics[width=\textwidth]{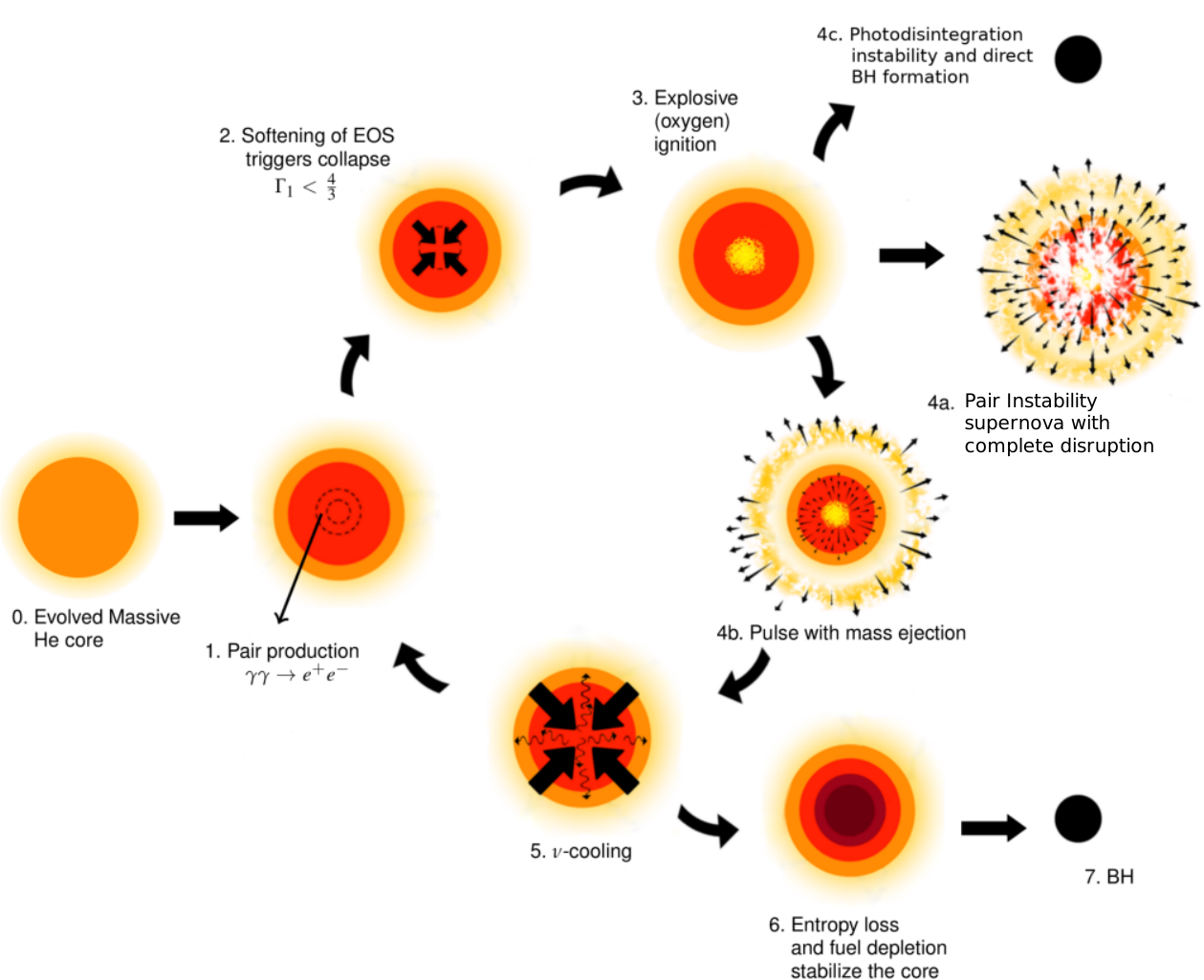}
  \caption{Evolution of a massive He core undergoing (pulsational) pair
    instability evolution. Three final outcomes are possible: full disruption without a compact remnant
(4a.), formation of a BH because of the photodisintegration instability (4c.),
or episodic mass loss (4b.) and final stabilization of the core, followed by a
regular core-collapse event.}
  \label{fig:cartoon}
\end{figure*}

Massive stars can have diverse final fates depending on the
structure of their core at the end of their evolution. This
diversity in terms of the kind of collapse (``electron-capture'',
iron-core collapse, pair-instability), whether it triggers an
explosion, albeit possibly weak, or not, and which remnant is left
behind is poorly understood as of yet. Typically, the core becomes dynamically unstable and starts
collapsing when no viable nuclear fuel is left. However, for very
massive radiation-pressure dominated stars, the core becomes
dynamically unstable while net energy generation by thermonuclear
reactions is still possible. 

Stars that end
their main sequence with a helium (He) core  exceeding
$M_\mathrm{He, init}\gtrsim80\,M_\odot$, which decreases to
$M_\mathrm{He}\simeq60\,M_\odot$ accounting for the wind mass loss, are predicted to end their evolution as
pair-instability supernovae \citep[PISN,][see also \Figref{fig:cartoon}]{fowler:64,rakavy:67}. They
evolve in hydrostatic equilibrium until they develop a
carbon-oxygen (CO) core. Soon after, the conversion of photons into
electron-positron (\pair{}) pairs (step 1 in \Figref{fig:cartoon})
causes a softening of the equation of state (EOS), initiating the
collapse of the star (step 2 in \Figref{fig:cartoon}). This increases the inner temperature until explosive
thermonuclear oxygen burning (step 3 in \Figref{fig:cartoon}) reverts the collapse and fully disrupts the star
\citep[e.g.,][step 4a in \Figref{fig:cartoon}]{barkat:67, fraley:68, kasen:11, yoshida:16, woosley:17, woosley:19}. Such stars do not
leave any compact remnant at the end of their evolution.

For initial
$M_\mathrm{He, init}\gtrsim200\,M_\odot$, decreasing to
$M_\mathrm{He}\simeq 125\,M_\odot$ after wind mass loss, stars also experience explosive thermonuclear oxygen
burning but, owing to energy loss as a result of the photo-disintegration of heavy nuclei, the
explosion is not energetic enough to reverse the collapse into an explosion and
disrupt the star, \citep[e.g.,][step 4c in \Figref{fig:cartoon}]{bond:84,fryer:01,heger:03}. In these
cases, the final fate is core collapse (CC), forming a massive black hole (BH). Therefore,
if these stellar explosions do occur in nature, a ``PISN black hole mass gap''
(also called ``second mass gap''\footnote{The ``first gap'' is the apparent
  lack of compact objects with masses between the maximum neutron star mass,
  $\max\{M_\mathrm{NS}\}\simeq 2\,M_\odot$ and the least massive BH known
  $\min\{M_\mathrm{BH}\}\simeq 5\,M_\odot$, (e.g., \citealt{farr:11}, but see also
  \citealt{wyrzykowski:16, wyrzykowski:19}).}) is expected between the most massive BH that can
be formed without encountering a PISN fate and the least massive BH formed because
of the photodisintegration instability.

The most massive BHs below the gap result from the evolution of He cores with
final masses just below $\sim$\,$60\,M_\odot$ \citep[e.g.,][]{yoon:12, woosley:17,
  farmer:19}. In these stars, the explosive burning of step 3 in
\Figref{fig:cartoon} releases less energy and
thus is only able to eject a fraction of the outer layers of the star. This
produces a mass-loss pulse 
(step 4b in \Figref{fig:cartoon}), without fully disrupting the star \citep{rakavy:67,
  fraley:68, woosley:02, woosley:07, woosley:17, woosley:19}. This phenomenon is the
lower-mass analog of a PISN, a pulsational pair-instability (PPI). The star
may undergo multiple such pulses until the combined effects of pulsational mass
loss, entropy loss to neutrinos (step 5 in \Figref{fig:cartoon}), and fuel consumption stabilizes the core
\citep[][]{woosley:17, marchant:19, farmer:19, leung:19}. Ultimately, this star is likely
to collapse to a BH, possibly with an associated supernova (SN), at
step 7 in \Figref{fig:cartoon}.

Given the impact on the distribution of BH masses \citep[][]{belczynski:16,
  woosley:17,marchant:19, stevenson:19}, the recent direct detection of
gravitational waves \citep{LVC:17, GWTC1} has revived the interest in PPI
evolution. Moreover, the follow-up of gravitational wave merger events is driving
large observational efforts in time-domain astronomy, with new and upcoming
facility such as the Zwicky Transient Factory \citep[][]{bellm:14}, Large
Synoptic Supernova Survey \linebreak \citep[][]{LSSThandbook}. When not following up
gravitational wave events, these instruments will perform surveys of various
depth and cadence which will soon unveil the variety of electromagnetic
transients possible in the sky. The James Webb Space Telescope will be able to
probe the transients expected at the death of the first stars in the
Universe, increasing the chance of a direct unambiguous detection of PISN or
PPI \citep[][]{whalen:13, regos:20}. Another potential piece of
  indirect evidence for the occurrence of PISNe is
a peculiar distribution of isotopes in their yields,
due to the neutron-poor type of nucleosynthesis
\citep[so called odd-even effect, e.g.,][]{woosley:02}. However,
the detection of this effect in the surface mass fractions of low
metallicity stars
remains debated \citep[e.g.,][]{aoki:14}.

Therefore, it is important to characterize the observable
characteristics of PPI evolution, that is, address the
question of what are the observable signatures of a pulse. Of particular
interest is the question of how much
mass do the pulses eject and at what velocity is it launched \citep[][]{leung:19}, or in other
words, what are the
circumstellar material (CSM) structures that this process can produce.

Previous studies from \cite{chatzopoulos:12, chatzopoulos:12b} investigated the
fate of (hydrogen-rich) stars with zero age main sequence (ZAMS) masses above
$40\,M_\odot$, with and without rotation during the pre-explosion evolution\footnote{The effect of
  rotation on the explosion dynamics has been investigated in
  \cite{glatzel:85, chatzopoulos:13}.}, and found PPI evolution in
the initial mass range $40-65\,M_\odot$ (for high rotation rates) and
$80-110\,M_\odot$ (without rotation). These ranges are also sensitive
to the details of the nuclear physics \citep[e.g.,][]{takahashi:18,farmer:19}.

\cite{woosley:17, woosley:19}, building up on previous work by
\cite{woosley:02, woosley:07}, presented the
first grids of stellar evolution calculations for a wide mass range
enclosing both PPI followed by a core collapse (PPI+CC) and PISN. 
The light curves of the former are expected to show a series of brightening
events as the individual pulses collide with each other \citep{woosley:17}, which has been proposed to explain the extremely luminous light curve of
SN2006gy \citep[][]{woosley:07}. More recently, \cite{arcavi:17, woosley:18}
also proposed PPI as a way to explain the peculiar photometric and spectroscopic
evolution of SN iPTF14hls, possibly coming from a merger progenitor
\citep[][]{vignagomez:19, spera:19}.

Although PISN need not be extremely luminous \citep[][]{woosley:17}, they are routinely considered in the
context of super-luminous supernovae 
\citep[e.g.,][]{gal-yam:09,chatzopoulos:13}. No
unambiguous identification of an astrophysical transient with a PISN
is available as of yet, however \cite{kozyreva:18} proposed OGLE14-073 as
a promising candidate. For models evolving through PPI before their final
collapse, recently claimed observational candidates are PTF12dam, a fast rising
type I super-luminous SNe modeled by \cite{tolstov:17} with a combination of
CSM interaction and radioactive decay; iPTF16eh, a type I
super-luminous SN showing signs of a shell of circumstellar
material through the detection of a light echo
\citep[][]{lunnan:18}; and SN2016iet, for which a dense, hydrogen (H)- and He-free CSM at
$10^{15}\,\mathrm{cm}$ of the star can be invoked to explain the light curve
\citep[][]{gomez:19}. Other potential candidates are type Ibn SNe showing
relatively narrow He lines, such as SN2006jc, whose progenitor was
observed to experience an outburst two years before the final
explosion \citep[][]{pastorello:07,foley:07}, and PS15dpn whose light
curve has also been modeled with a combination of CSM interaction and
$^{56}\mathrm{Ni}$ decay by \cite{wang:19}.

Here we calculate the detailed evolution of massive He
cores to characterize jointly the effect that PPI evolution has on the final BH
masses and on the circumstellar material (CSM) structure. Our stellar evolution
models can provide input for the hydrodynamical evolution of the CSM,
which can become visible because of collisions between shells of
CSM ejected at different times
\citep[e.g.,][]{woosley:07,woosley:17}, or possibly if the BH form with an accompanied explosion. We also
provide (\emph{i}) a criterion to determine which He cores encounter a \emph{global}
instability, resulting in PPI-driven mass loss and BH formation,
and which He cores instead are fully disrupted in a PISN and (\emph{ii}) the bulk properties of the pulses
and their distribution as a function of mass.

In \Secref{sec:methods} we describe our calculations, 
before giving an overview of the
evolutionary outcome of our models in \Secref{sec:overview} and of the resulting
BH masses in \Secref{sec:bh_masses}. We focus on the PPI models in
\Secref{sec:pulses_def}, where we describe three physically motivated possible
definitions of a ``pulse''. While the basic ideas on how the
evolution of these models proceeds are well established from the theoretical side, there is some ambiguity
in the literature on what is called a pulse. We discuss the CSM that our models
can produce with a toy-model assuming propagation of the
ejecta at constant velocity in \Secref{sec:grid_ejecta}, and provide input files for a more
sophisticated modeling of the CSM at
\href{https://doi.org/10.5281/zenodo.3406356}{https://doi.org/10.5281/zenodo.3406356}. In 
\Secref{sec:explodability} we discuss whether the final core collapse after the
PPI evolution would produce an associated SN explosion, which would generate
ejecta to interact with the previously ejected stellar layers. We compare
our results to a few observational transients that have been interpreted as
pulsational pair-instability events in \Secref{sec:obs_SNe}. We define a
criterion to distinguish pulsational evolution from full disruption without
going through the hydrodynamic calculations in \Secref{sec:caveats}, before highlighting the main limitations of this study. \Secref{sec:conclusions} summarizes our main conclusions.
Appendix~\ref{sec:res_study} presents a resolution study of one of our models,
and Appendix~\ref{sec:hrichcomp} compares the evolution of a naked He core to a
H rich star with a similar He core mass.

\section{Pulsational pair-instability evolution with MESA}
\label{sec:methods}

We model the evolution of bare He cores because stars massive enough to encounter the
PPI are likely to have lost their H-rich envelop beforehand. This could happen either because of the presence of a
binary companion
\citep[e.g.,][]{kippenhahn:67},
strong wind mass loss \citep[e.g.,][]{vink:05}, or because of rotational mixing
preventing the formation of a core-envelope structure \citep[][]{maeder:00, yoon:06,
  demink:09, mandel:16b, marchant:16}.

Another way to form very massive stars which are expected to produce
the most massive (stellar mass) BHs
is through runaway collisions in a dynamically excited environment
\citep[e.g.,][]{vandenheuvel:13}, or binary mergers
\citep[e.g.,][]{demink:14, vignagomez:19}. Either might result in the loss from the system of some H-rich material. Even if 
a binary merges before the onset of pulsations and retains a significant amount of
H \citep[e.g.,][]{vignagomez:19}, the merger may well undergo asteroseismologic (non-PPI) pulsations enhancing wind mass
loss removing of the remaining envelope
\citep[][]{moriya:15}. Finally, any remaining H-rich envelope is likely
to be loosely bound and easily removed during the first PPI pulse
\citep[][see also Appendix~\ref{sec:hrichcomp}]{fraley:68, leung:19}. 

We employ the open-source stellar evolution code \texttt{MESA}
\citep[release 11\,701,][]{paxton:11,paxton:13,paxton:15,paxton:18,
  paxton:19} to evolve a grid of He stars in the mass range
$35\,M_\odot\lesssim M_\mathrm{He,init}\lesssim 250\,M_\odot$. Throughout
this study, we define the He core mass as the total mass of our
models, and the CO core boundary as the outermost
location where the mass fraction of $^{4}\mathrm{He}$ drops below
0.01. We adopt an initial metallicity of $Z=0.001$, and we also ran a
limited sample of models with $Z=0.00198$ \citep[similar to the value
quoted for SN2016iet,][]{gomez:19}. Both these values are below the
upper limit for the occurrence of PISN of $Z_\odot/3\simeq0.006$
obtained from single star models \citep{langer:07}. Pair-instability
evolution might even occur at higher metallicity because of late
stellar mergers in a binary \citep[][]{vignagomez:19}, or if magnetic
fields funnel the mass lost to winds back to the star
\citep[][]{georgy:17}. We do not study here the impact of binarity,
rotation, and magnetic fields.

We include wind mass loss as in \cite{marchant:19} and in the fiducial
model of \cite{farmer:19}, that is we use the rate from
\cite{hamann:95,hamann:98} reduced by a factor of 10 to account for
wind clumpiness. For effective temperatures
$T_\mathrm{eff}<23\,300$\,K, which can be achieved in between pulses
due to the expansion of the star, we employ the maximum between the
\cite{vink:00,vink:01} and \cite{nieuwenhuijzen:90} wind mass loss
rates. We turn off wind mass loss during the (physically brief)
dynamical phases of evolution: PPI-driven dynamical mass ejections are
the only source of mass loss in these phases.  Uncertainties in the
wind mass loss rate can have an impact on the core structure
\citep[][]{renzo:17}. The main effect of varying the wind
algorithm in our H-free models is to change
  the mapping of the initial He core mass $M_\mathrm{He,init}$ to the
  He core mass at the onset of the
  pulses. Stronger wind mass loss would inevitably reduce the mass of
  PPI-produced CSM by removing the mass before the
  instability. Moreover, the wind velocity might differ from the
  ejection velocity in pulses, thus resulting in a different CSM
  density profile.  In \cite{farmer:19} we explored the uncertainties in
  the wind mass loss rates, varying their functional form (including
  empirically determined rates from \citealt{nugis:00} and
  \citealt{tramper:16}), and efficiency factors. We found that these
  uncertainties do not significantly influence the range of possible
  BH remnant masses and the PPI+CC properties when expressed as a
  function of the carbon-oxygen core mass. For a recent
  comparison of wind mass loss rates for Wolf-Rayet stars, we refer
  the interested reader to \cite{yoon:17} and to \cite{woosley:19} for a grid of models
  spanning the PPI-regime.

To follow the dynamical evolution of the pulses when they occur, we
use \texttt{MESA}'s Riemann HLLC solver \citep[][]{toro:94, paxton:18}. We
determine the dynamical stability of the star based on the adiabatic index
$\Gamma_1=\partial \log(P)/\partial \log(\rho)|_s$. Regions of the star with
$\Gamma_1>4/3$ are formally stable
\citep[e.g.,][]{kippenhahn:13}. However, this is a local
quantity. To create a global metric descriptive of the entire star, we follow
\citealt{stothers:99} in defining a volumetric pressure-weighted 
average adiabatic index 
\begin{equation}
  \label{eq:pulse_criterion}
  \langle\Gamma_1\rangle \udef \frac{\int \Gamma_1P\,d^3r}{\int P\,d^3r} \equiv \frac{\int \Gamma_1 \frac{P}{\rho}\,dm}{\int
    \frac{P}{\rho}\,dm} \ \ ,
\end{equation}
where $P$, $\rho$ are the local pressure and density, and we used the
continuity equation to transform the volumetric integral into an
integral over the mass domain. Weighting the \emph{local} $\Gamma_1$
with $P$ makes the average $\langle\Gamma_1\rangle$ a dynamically
relevant quantity, and guarantees that the inner regions contribute
more to the average. Whenever $\langle\Gamma_1\rangle =4/3+0.01$,
i.e., slightly before the stellar structure becomes formally unstable,
we switch to a hydrodynamical treatment of the evolution and turn off
the stellar winds (see also \citealt{marchant:19}).

After a pulse, if the
internal structure of the star meets the criteria specified in
\cite{marchant:19} to conservatively ensure hydrostatic equilibrium has been recovered, we excise the material moving
faster than the local escape velocity and create a new star with the entropy,
chemical composition, and mass of the layers remaining bound. Even for
nonpulsating models, we turn on the hydrodynamics to follow the onset of
core-collapse, when the core temperature rises above $T_c\gtrsim10^{9.6}$\,K.

We adopt a 22-isotope nuclear reaction network
(\texttt{approx21\_plus\_co56.net}), which is sufficient to trace the 
energy output during the relevant burning phases but not the detailed
nucleosynthesis \citep[e.g.,][]{farmer:16, farmer:19}.

We assess convective stability using the Ledoux criterion, and adopt a
mixing length parameter of $\alpha_\mathrm{MLT}=2.0$. We consider
semi-convective mixing with an efficiency $\alpha_{s}=1.0$, whilst
neglecting thermohaline mixing. We assume an exponential
undershooting and overshooting with parameters\footnote{cf. Equation 2 in
  \citealt{paxton:11} and the \texttt{MESA} documentation for the
  definition of \texttt{f} and \texttt{f$_0$}.}
(\texttt{f},\texttt{f$_0$})=(0.01, 0.005) for all convective
regions. We follow the approach of \cite{marchant:19} based on
\cite{arnett:69} for the time-dependence of the convective
velocity. This is required to compute dynamical phases of the
evolution with timesteps shorter than the convective turnover
timescale \citep{renzo:20:conv_PPI}.

We stop our evolution either at the onset of CC or at the onset of a
PISN. We define the former as when the infall velocity anywhere in the
model exceeds $1000\,\mathrm{km\ s^{-1}}$ \citep[][]{woosley:02}. For
the latter we check that the total energy (including the kinetic term)
of the star is positive and that the minimum radial velocity is
non-negative. These conditions
guarantee that the star is unbound and there is an outflow of matter.

We define the CO core mass ($M_\mathrm{CO}$) of our models as the
maximum mass coordinate where the mass fraction of He is lower than
$Y<0.01$ at He core depletion, i.e.~when the central mass fraction
  of He reaches $X_c(^{4}\mathrm{He})<10^{-5}$. This
allows us to characterize our models with one single CO core mass,
while the actual amount of CO-rich mass might change because of
the evolution. We define the iron core mass ($M_\mathrm{Fe}$) as the outermost
mass location where the abundance of $^{28}\mathrm{Si}$ is lower than
0.01 and the abundance of elements with mass number $A>46$ is greater
than 0.1. We use discuss $M_\mathrm{Fe}$ only at the onset of CC. We refer the reader to
Appendix~\ref{sec:res_study} for the description of the numerical
resolution and a quantitative assessment of its impact on our results.

The input files (\texttt{inlist}s) and customized routines added to the code
(\texttt{run\_star\_extras.f}) needed to reproduce our results are available at
\href{https://doi.org/10.5281/zenodo.3406356}{https://doi.org/10.5281/zenodo.3406356}. We also provide our numerical results for the evolution and final structure of each of our models, including a
customized output file storing averaged information for each layer moving beyond
its local escape velocity during PPI-driven mass loss episodes. The
  possibility of fallback is neglected in these files, even though our \texttt{MESA}
  models allow for it. Such files can
be used as inputs for hydrodynamic studies of the CSM structure
produced by these stellar models.

\begin{figure*}[hbtp]
  \centering
  \includegraphics[width=\textwidth]{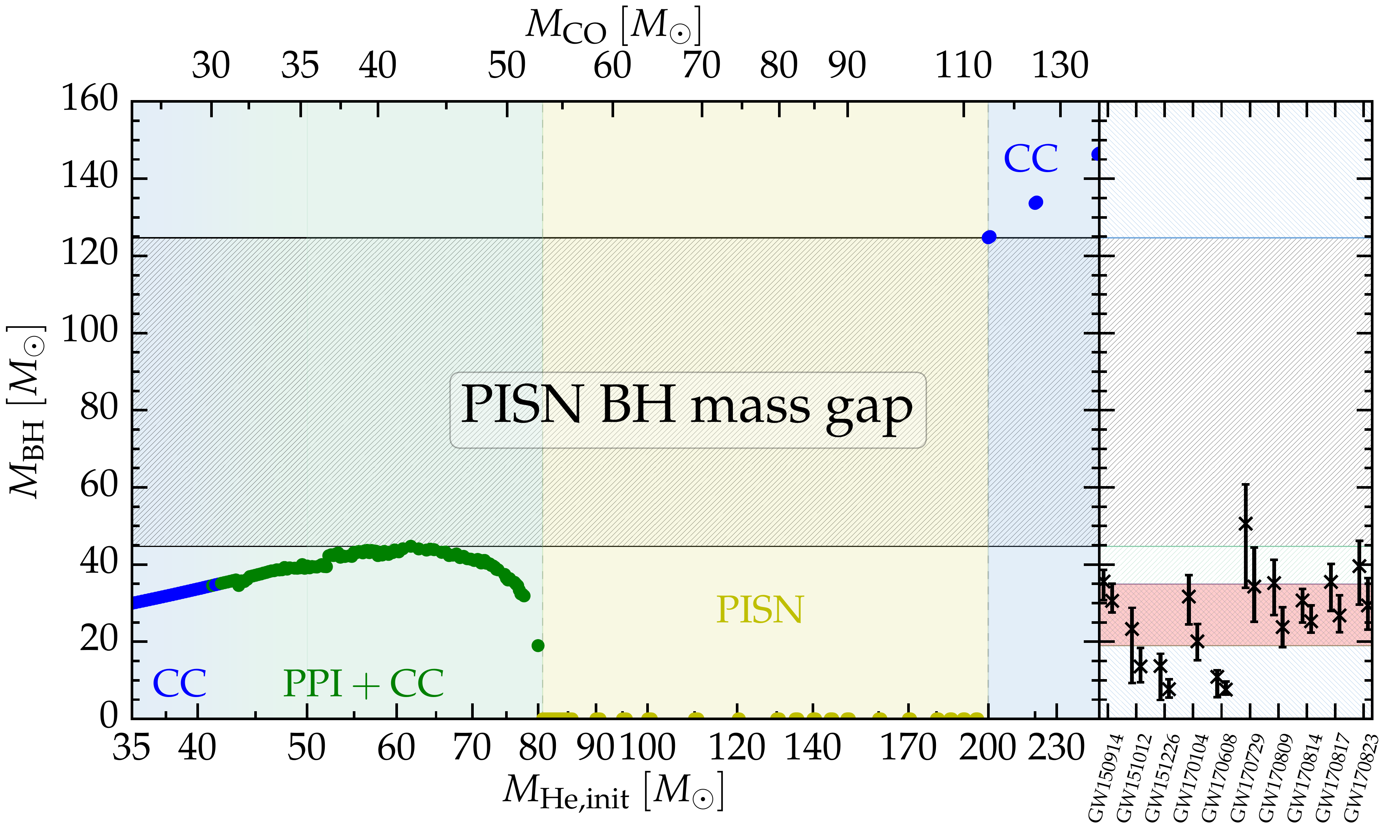}
  \caption{Final BH masses as a function of the initial He core
    mass. The scale in the horizontal direction is logarithmic. The colors in
    the background indicate the approximate range for each evolutionary path,
    see also \secref{sec:overview}. The right
    panel shows the masses inferred from the first ten binary BH mergers detected by
    LIGO/Virgo, with a red shade to emphasize the overlap between PPI and CC,
    and green and blue hatches to indicate the fate of the progenitor in
    different BH mass ranges.}
  \label{fig:mhe_mbh}
\end{figure*}

\section{Overview of the evolution of the progenitors}
\label{sec:overview}

\figref{fig:mhe_mbh} shows the BH masses resulting from our grid as a function of
the initial He core mass ($M_\mathrm{He,init}$, bottom axis) and approximate
maximum CO core mass  reached during the evolution ($M_\mathrm{CO}$, top axis). Both can
 decrease because of PPI mass-loss episodes toward the end of the evolution. We
 estimate the BH mass as the mass coordinate where the binding
energy of the collapsing star reaches $10^{48}\,\mathrm{ergs}$, to allow for the
possibility of mass loss during the CC from, either a weak
explosion \citep[][]{ott:18, chan:20}, or ejection of a fraction of
the envelope due to neutrino losses\linebreak \citep[][]{nadezhin:80, lovegrove:13}. This estimate
is typically within a few $0.01\,M_\odot$ of the total final mass of the He star. We do
not account for other energy loss terms during the core collapse, such
as neutrinos themselves which might carry away (part of) the core binding energy. This effect is
typically estimated to be $\lesssim$\,10\% of the
precollapse core rest mass energy \citep[e.g.,][]{oconnor:11,belczynski:16,
  spera:17},
and can shift our BH mass estimates further down.

The colored background in the left panel of \Figref{fig:mhe_mbh} indicates approximately the
evolutionary path for the corresponding mass range. The four possibilities
are summarized as follows, in order of increasing initial He core mass:

\paragraph{\bf CC:} Relatively low mass He cores end their lives
in a core collapse (CC, blue on the left of \Figref{fig:mhe_mbh}) event without losing mass to pair-production
driven pulses. For these models, the layers which are unstable to pair production
(if any) are not massive enough to cause an episode of mass ejection. In this
mass range, the outcome of core-collapse is most likely BH formation, possibly
associated with a weak SN with large fallback
\citep[][]{ott:18,kuroda:18, chan:18, chan:20}. We
return on the ``explodability'' of our grid of models in \Secref{sec:explodability}.

\paragraph{\bf PPI+CC:} With increasing $M_\mathrm{He,init}$, the pair instability
becomes progressively more violent. The energy release by thermonuclear explosions causes
significant radial expansion. Increasing further in mass, models experience
one or more mass loss episodes, before the core is stabilized by the consumption
of fuel and entropy losses to neutrinos, and the stars finally collapse (PPI+CC,
green in \Figref{fig:mhe_mbh}).
\paragraph{\bf PISN:} For $80\,M_\odot \lesssim M_\mathrm{He,init}\lesssim200\,M_\odot$, our models
  are completely disrupted in a PISN, and produce no remnant (yellow vertical
  area in \Figref{fig:mhe_mbh}). Our lowest mass model going PISN and leaving no remnant has
$M_\mathrm{He,init}=80.75\,M_\odot$, corresponding to a maximum CO core mass of
$\sim$\,$55\,M_\odot$ \citep[see also][]{farmer:19}.

\paragraph{\bf CC:} For extremely massive cores, $M_\mathrm{He,init}\gtrsim200\,M_\odot$,
 the energy release by the explosive thermonuclear burning triggered by the
 pair instability is insufficient to fully disrupt the
 star. This happens because most of that energy is used to photodisintegrate the
 nuclear ashes and lost to neutrinos, instead of becoming kinetic
 energy of the stellar gas \citep{bond:84,fryer:01}. Therefore, models above a certain threshold reach CC without any PPI-driven
 mass loss (blue area on the right of the panel of \Figref{fig:mhe_mbh}).\\
 
In \Figref{fig:mhe_mbh} the transition between the CC and PPI+CC is smooth, and
we have avoided quantifying the boundary between the low mass CC and PPI+CC
because of the subtleties in the definition of ``pulse''. We outline three
physically motivated definitions, each one shifting the CC/PPI+CC
boundary, in \Secref{sec:pulses_def}.

\section{Resulting BH masses}
\label{sec:bh_masses}

The PISN BH mass gap is denoted by the hatched region in
the left panel of \Figref{fig:mhe_mbh}. The lower and upper edge of the gap can be read from the
y-axis.  With our numerical setup, we find a maximum BH mass below the
PISN gap of $\max\{M_\mathrm{BH}\}\simeq45\,M_\odot$, in good agreement with
the lower boundary of the gap from previous studies \cite{woosley:17, marchant:19, leung:19, woosley:19, farmer:19}. At the
upper-end, the PISN BH mass gap is closed by the
photodisintegration instability causing the direct collapse of an initially
$M_\mathrm{He,init}=200\,M_\odot$ He core, which builds up a
CO core of $M_\mathrm{CO}\simeq 114\,M_\odot$ and eventually forms to
a BH of $125\,M_\odot$. This value corresponds very closely to the
final He core mass of this model, after wind mass loss. This upper boundary too is in good agreement with the
results from \cite{woosley:02,woosley:17}, although it is sensitive to the
metallicity and uncertainties in the wind mass loss rate.

We do not expect these boundaries would have varied if our models had a H-rich
envelope \citep[e.g.,][]{woosley:17}, especially if the progenitor
stars evolve in close binaries which can remove the H-rich envelope
long before the PPI. The combination of the mass loss
\citep[][]{ott:18, kuroda:18, chan:18, chan:20} and energy loss to neutrinos at BH
formation \citep[e.g.,][]{coughlin:18} should be sufficient to unbind
any the residual H envelope, unless the progenitor is a blue super giant with a large binding
energy of the envelope (exceeding $\sim$\,$10^{48}\,\mathrm{erg}$,
\citealt{lovegrove:13}). Such blue supergiant pre-PPI structures might arise
from binary mergers \citep[e.g.,][for a population synthesis
study]{spera:19}. However, the stellar structure calculations for the
merger of two post-main-sequence stars
from \cite{vignagomez:19} show
extended convective envelopes at the onset of the instability, which support our
expectation that the envelope would easily be shed at the onset of the
instability (see also Appendix~\ref{sec:hrichcomp}). Whether the H-rich envelope
can contribute to the BH mass or not deserves further investigation.

The right panel of \Figref{fig:mhe_mbh}
shows for comparison the individual BH masses of the binary BH mergers detected
to date by LIGO/Virgo\footnote{Other events have since been reported by an
  independent analysis of the first two observing runs, see \cite{zackay:19} and
  references therein.}, with the 90\% confidence level uncertainty ranges. The masses of the two BHs in a merger event are not direct
observables, they are instead
inferred from the chirp mass and total mass of the binary. The color
of the hatching in the right panel indicates
the possible progenitor evolution (see also \Secref{sec:overview}):
the red area emphasize the range of BH masses that can be obtained by CC of a lower
mass model, or by severe PPI mass loss of the most massive PPI+CC
models. Its extent to the lower BH masses is somewhat dependent on the
resolution in $M_\mathrm{He,init}$ of our grid. However, in \cite{marchant:19}
we showed that the minimum BH mass that can be obtained by PPI+CC
evolution is about 10\,$M_\odot$ because of the production of
radioactive material that can unbind cores that recover hydrostatic
equilibrium of lower masses. Most BH progenitors
for the gravitational wave mergers events detected to date are compatible
with encountering the PPI, although we do not expect most of them to have gone
through this evolution because progenitors with sufficient mass are disfavored by the
initial mass function.

\section{The physics of pulses: cores, radii, and mass ejections}
\label{sec:pulses_def}

While the nuclear and thermal processes
governing the evolution of a star through pair instability are well understood, the
characterization of the observable properties of such events are not yet as
clear. The main reason for this is that stars do not react ``elastically'' to
the pair instability: instead the nuclear binding energy released by burning
episodes at each pulse is stored and re-distributed throughout the stellar
structure, and there is not a one-to-one correspondence between what happens in
the core and what can be observed at the surface.

To clarify the distinction between core behavior and observable properties from
the outermost layers of the star, in this section we describe the physical
processes that can be used to give three different physically-motivated
definitions of ``pulse'', and illustrate them with an example $M_\mathrm{He,init}=50\,M_\odot$ He core (for which we
also present a resolution study in Appendix~\ref{sec:res_study}). These
definitions do not cover all the possible ways in which pulses can be defined
and counted. For example, \citealt{woosley:07,woosley:17} uses the core temperature $T_c$
while in \citealt{marchant:19} we adopted a criterion based on the maximum velocity in the
stellar interior.

\subsection{Thermonuclear ignition}
\label{sec:thermo_def}

Historically, studies on pair-instability evolution have focused on the
core of stars. Indeed, the region that becomes unstable because of the \emph{runaway}
production of $e^\pm$ is typically deep in the star, and the subsequent
evolution is driven by the explosive burning of oxygen and heavier fuel 
\citep[e.g.,][]{barkat:67, rakavy:67,fraley:68, woosley:07, woosley:17, marchant:19,
  leung:19}. This allows for a definition of a pulse based on the behavior of
the deep interior of the star.

\begin{figure}[htbp]
  \centering
  \includegraphics[width=0.5\textwidth]{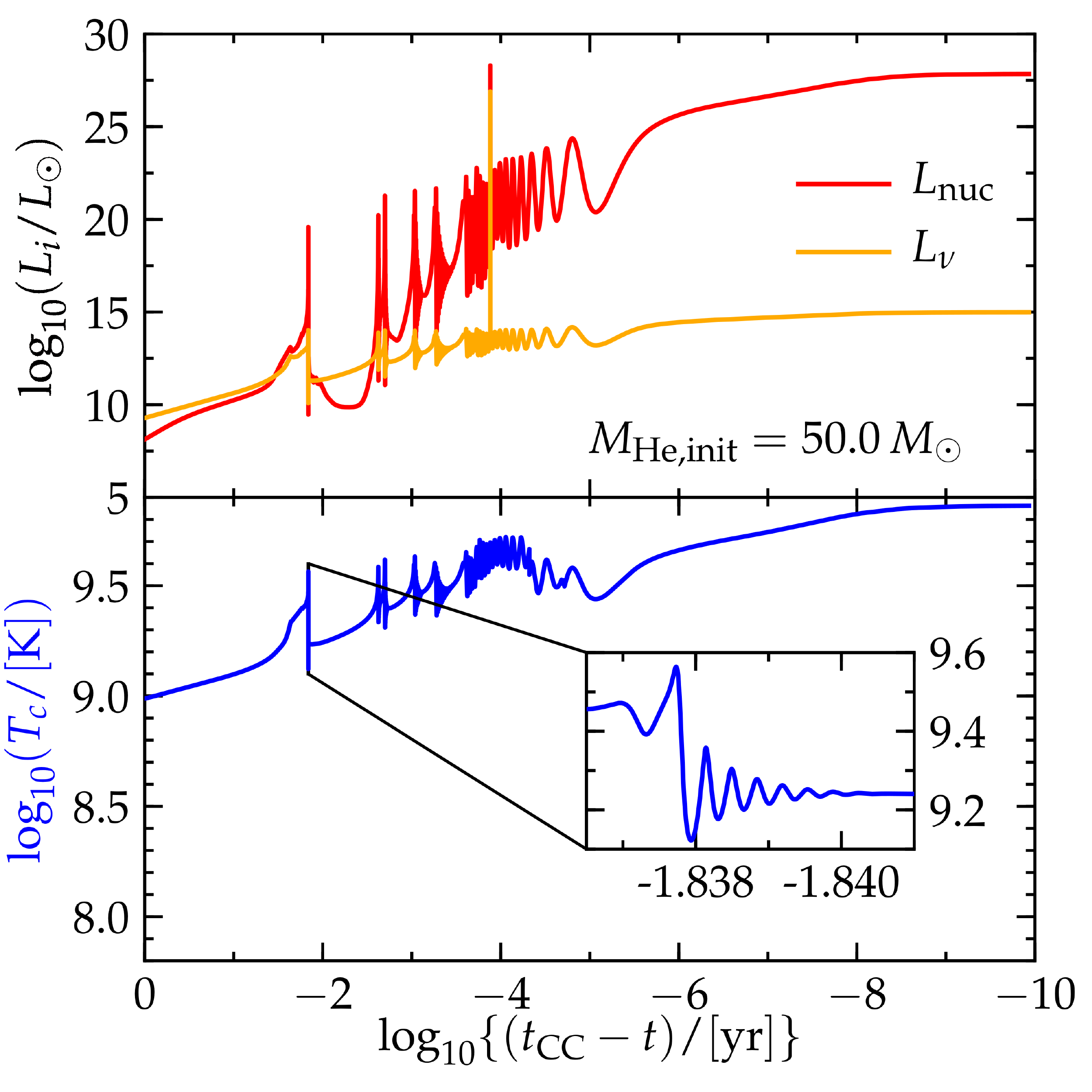}
  \caption{Last year of evolution of the central temperature (blue, bottom panel), nuclear
    (red) and neutrino (orange) luminosity (top panel) for a $50\,M_\odot$ He
    core. The inset in the bottom panel shows the readjustment of the core to
    hydrostatic equilibrium after the first pulse, which we resolve even
    if this behavior is likely to be influenced by the imposed
    spherical geometry.}
  \label{fig:TcL}
\end{figure}

\Figref{fig:TcL} shows the evolution of the central
temperature ($T_c$, bottom panel) and nuclear and neutrino luminosity
($L_\mathrm{nuc}$ and $L_\nu$ respectively, top
panel) during the last year before CC for a
$M_\mathrm{He,init}=50\,M_\odot$ He core. The horizontal axis shows the time to the onset of CC
on a reversed logarithmic scale.

During the previous evolution (not shown), this model 
is thermally and dynamically stable. At
$\log_{10}\{(t_\mathrm{CC}-t)/[\mathrm{yr}]\}\simeq2$, the core becomes thermally unstable
because of the softening of the EOS, causing the
collapse of the core and
rise in $T_c$ allowing for
the explosive ignition of fuel. The latter can be seen as spikes in the nuclear
luminosity $L_\mathrm{nuc}$. The thermonuclear release of energy expands the core, cooling
it adiabatically and causing a temperature drop. Eventually, the core is
stabilized by the loss of entropy to neutrinos and the burning of nuclear fuel, and it ends
its life steadily increasing its core temperature until
the onset of CC.

If we define PPI
pulses based on core temperature spikes, or equivalently spikes in nuclear and
neutrino luminosity, then the lowest mass He core showing hints of pulsational behavior is
$M_\mathrm{He,init}\simeq37.5\,M_\odot$, corresponding to a final
$M_\mathrm{CO}\simeq28\,M_\odot$. Even if the local adiabatic index $\Gamma_1<4/3$ somewhere
in this model, the volumetric pressure-weighted averaged adiabatic index
is always $\langle\Gamma_1\rangle>4/3$ for the entire evolution. The thermonuclear
ignition in the core never results in a \emph{global} instability of the star.
We find that $\langle\Gamma_1\rangle $ crosses the stability threshold of 4/3 at some point in the evolution only for $M_\mathrm{He,init}>40.5\,M_\odot$.

The inset plot in the bottom panel of
\Figref{fig:TcL} shows that in a one-dimensional spherical \texttt{MESA} model the core
``bounces'' off itself, which was already noted in \cite{paxton:18}. These
readjustments of the core cause secondary burning episodes
\citep{renzo:20:conv_PPI} which can
release further energy and introduce complications in counting the $T_c$
spikes (see also \citealt{marchant:19}). While we
resolve in time these bounces by taking timesteps shorter than the dynamical
timescale of the core, it is likely that multidimensional effects
and/or off-center energy release would affect
them significantly. Even counting the core oscillations as one individual pulse,
our $M_\mathrm{He, init}=50\,M_\odot$ model exhibits tens of
thermonuclear-ignition pulses.

This behavior of the core of very massive stars encountering the pair instability is well
established \citep[e.g.,][]{barkat:67, woosley:17, marchant:19, leung:19}.
However, since these processes happens deep inside the optically thick layers of
the star, their only direct observable is the rapid variation of orders of
magnitude of the neutrino luminosity during each
pulse (\citealt{fryer:01}, and possibly during the post-pulse bounces). However, because
of the rarity of such massive stars in the local Universe, such variations in
the neutrino luminosity are unlikely to be easily observed.

\begin{figure}[bp]
  \centering
  \includegraphics[width=0.5\textwidth]{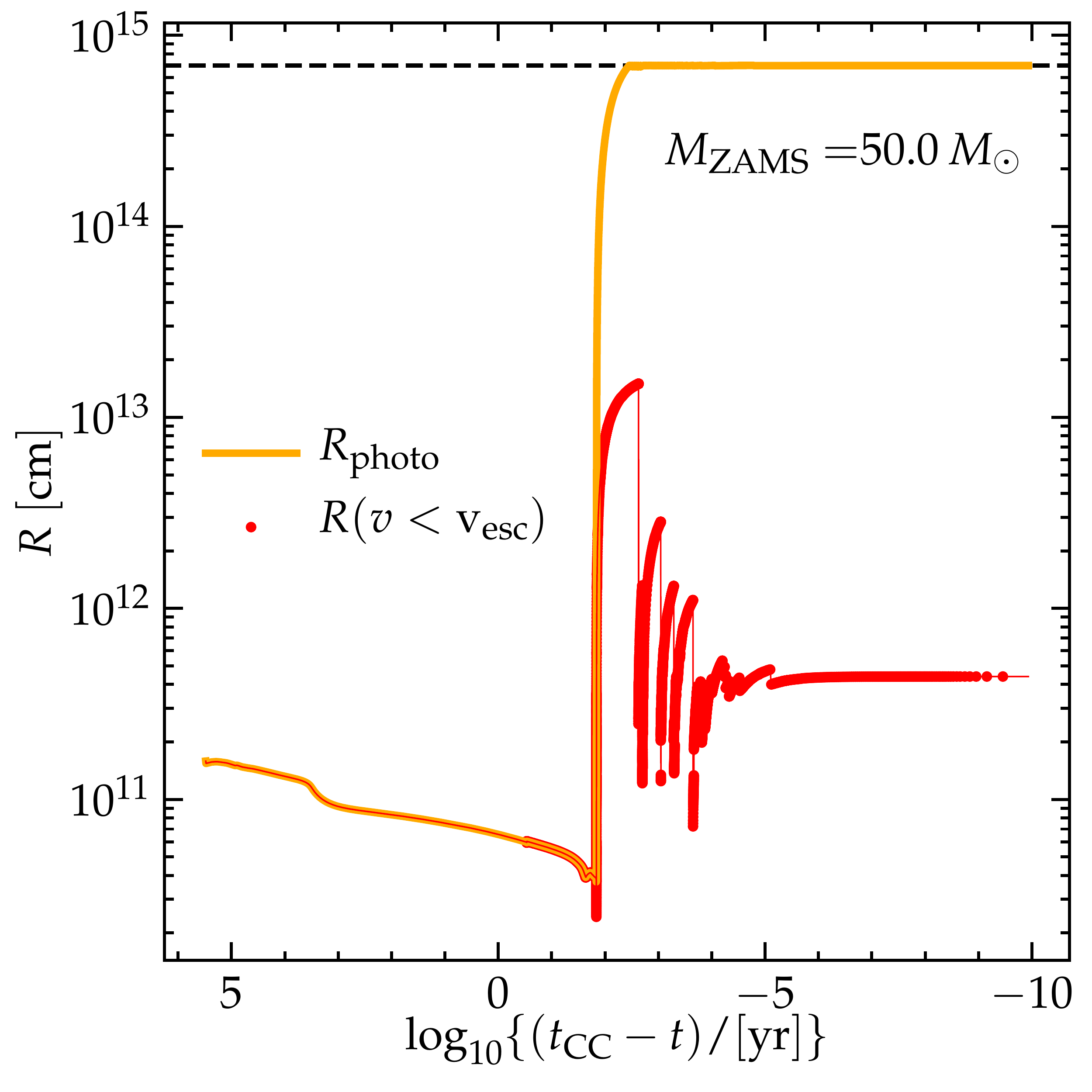}
  \caption{Radial evolution of a $M_\mathrm{init}=50\,M_\odot$ He core through PPI pulses. The radius of
    the bound material (red, plotted only during dynamical phases) oscillates
    because of the PPI. The photospheric radius (orange) reaches $10^4\,R_\odot$
    (dashed line) during the dynamical phase of evolution, i.e., the location at
    optical depth 2/3 is in the material already ejected. }
  \label{fig:R_t}
\end{figure}

\subsection{Radial expansion}
\label{sec:radial_def}

In \Secref{sec:thermo_def} we discussed a definition of a PPI pulse
based on the thermonuclear behavior deep inside the core. However, for
$M_\mathrm{He,init}>41\,M_\odot$, the nuclear binding energy released
by the thermonuclear explosions deep down can have an observable
impact on the surface of the star. We emphasize that either because of
other evolutionary processes, or because of a previous PPI mass-loss
episode, the surface can be a He rich layer for these
stars. Therefore, we can use our He core models to define a PPI pulse
based on surface properties, assuming the H-rich layers have been lost
before.

The thermonuclear burning 
injects energy into the core and drives a
pulse wave, which propagates down the decreasing density profile of the star and
eventually steepens into a shock. The core, post-explosion, readjusts and can contribute to
driving secondary shocks. There is a small range in mass, $41\,M_\odot\lesssim
M_\mathrm{He,init}\lesssim 42\,M_\odot$ in which these shocks, which can often catch up
with each other below the stellar surface, are not energetic enough to dynamically unbind any
significant amount of matter (see also \Secref{sec:mejection_def}).
Nevertheless, even in this mass range, they produce a potentially observable radial expansion of
the star. For models more massive than $M_\mathrm{He,init}\gtrsim42\,M_\odot$, the energy
released in the thermonuclear explosion also cause the ejection of material (see \Secref{sec:mejection_def}). 

\Figref{fig:R_t} shows the radial evolution of our $50\,M_\odot$
example. The orange line shows the photospheric radius (defined as the location
where the optical depth is 2/3). For most of the evolution in hydrostatic equilibrium (until
$t_\mathrm{CC}-t\simeq10^{-2}$\,years), the stellar radius is on the order of the
solar radius ($R_\odot\simeq6.9\cdot10^{10}\,\mathrm{cm}$) or less. As the star contracts
and approaches the instability, we switch to the HLLC solver at around
$t_\mathrm{CC}-t\lesssim 1$\, year, when the thicker red line appears in
\Figref{fig:R_t}. This line shows the radius of the bound material $R(v<v_\mathrm{esc})$
and is plotted only when the hydrodynamics is on.

We follow the thermal contraction of the star due to the pair instability, and
at $t_\mathrm{CC}-t\simeq 10^{-2}$\,years a shock wave propagating from the
core causes a radial expansion by two orders of magnitude on a dynamical timescale. The material remaining bound to
the star extends to $\sim$\,$10^{13}$\,cm. For our $M_\mathrm{He, init}=50\,M_\odot$ model, the pulse
also ejects matter and the ejected layer extends beyond
$10^{14}$\,cm. For numerical stability reasons we cap the radii at
$10^4\,R_\odot\simeq6.9\cdot10^{14}\,\mathrm{cm}$ (horizontal dashed line in \Figref{fig:R_t}), and treat this limit as an open boundary\footnote{In none
  of our models is such an upper limit in radius reached by the bound material}. We
discuss the ejected matter more extensively in \Secref{sec:mejection_def}.

The ejected layer can obscure the bound surface of the star: the
location where the optical depth is 2/3 extends all the way to the
outermost layers of our Lagrangian mesh. However, we emphasize that
the structure of the material moving faster than the escape velocity
should be recomputed accounting for radiative losses for a better
determination of the photosphere. It is possible that such stars would
exhibit large radius differences at different wavelengths, with some
that might even appear red during their maximal radial expansion.  For
this particular model, the photospheric radius does not have time to
recover its prepulse value, since the radial expansion only started
days before the final core-collapse. More massive models have more
violent pulses that drive the inner core farther out of thermal
equilibrium and for which it takes longer to recover the condition for
further (explosive or stable) nuclear burning (see also
\Secref{sec:grid_ejecta}): this can give time to the photospheric
radius to decrease again.

The bound radius instead experiences large oscillations between $10^{11}$\,cm
and $10^{13}$\,cm (the maximal
expansion reached initially). In this case,
these might not be directly obervable since they are embedded within the
pseudo photosphere of the ejecta. Models more massive than our example might
have rather long lived phases with large radii, which might have
implications for binary interactions \citep[e.g.,][]{marchant:19} and wind mass
loss physics.

\subsection{Ejection of material}
\label{sec:mejection_def}

\begin{figure}[tbp]
  \centering
  \includegraphics[width=0.5\textwidth]{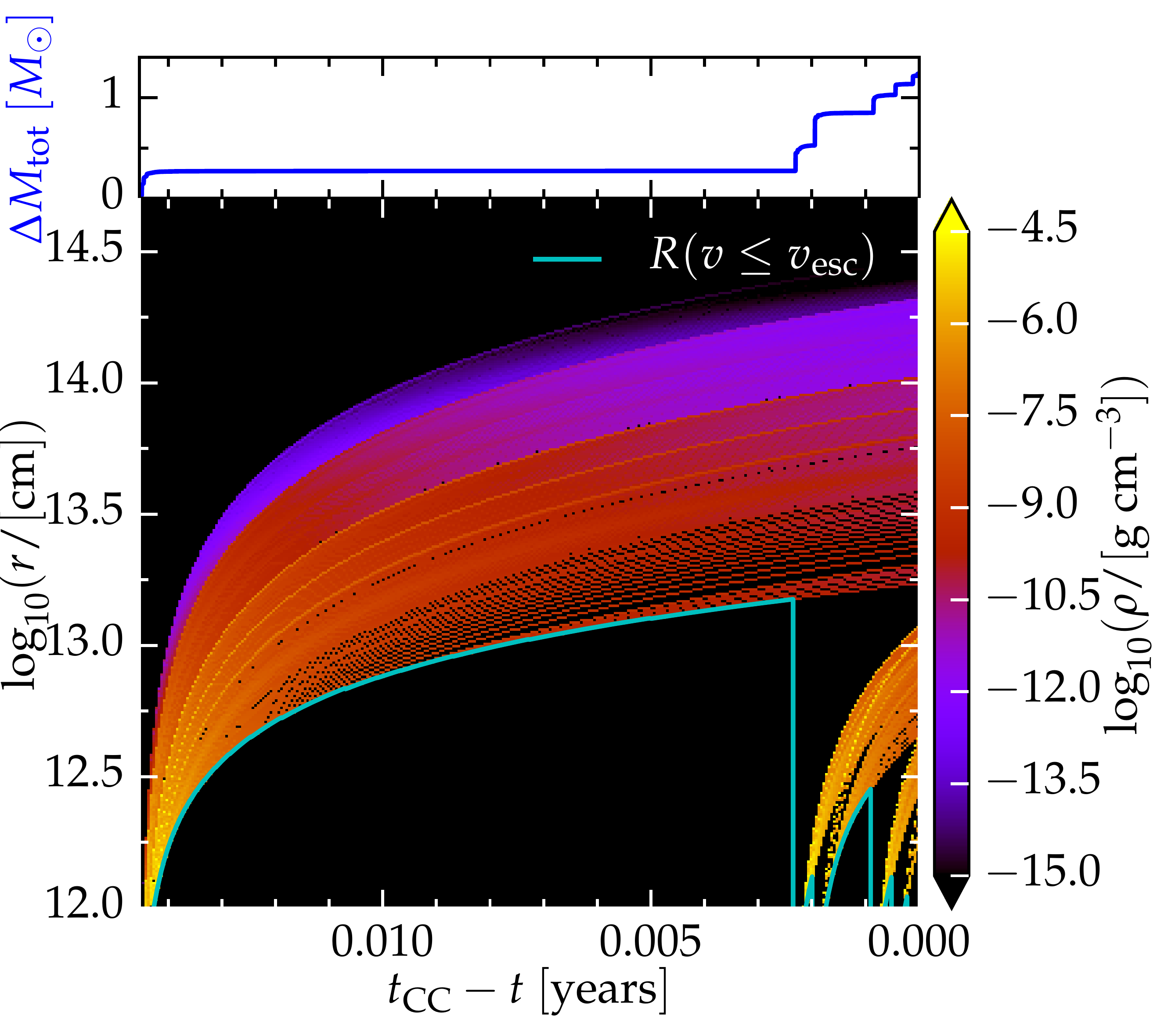}
  \caption{Space-time diagram for propagation of the PPI ejecta of a
    $50\,M_\odot$ He core. The
    color indicates the density, assuming radial expansion at constant velocity.
    The top panel indicates the cumulative amout of ejected at
    velocities larger than the escape velocity through pulses
    only (i.e., excluding the
    wind mass loss). The cyan curve shows the radius of the material
    instantaneously bound (cf.~the red curve in \Figref{fig:R_t}).}
  \label{fig:sptime}
\end{figure}

Although the two definitions of a pulse we introduced in the previous subsections (based
on the core explosive behavior and on the radial expansion, respectively) might
possibly give observable ``pulses'', the processes they are based on do not leave a direct imprint on the CSM
structure, nor on the remnant BH mass. Observational confirmation of the occurrence in
nature of PPI+CC evolution (and possibly PISN) is most likely to come
from either observations of
transients which can probe the CSM around the exploding star and/or the
distribution of BH masses probed through gravitational waves.
It is therefore worth giving a definition of pulse based on the
nonterminal ejection of
material from the stars: the ejecta carry away mass, decreasing the final BH mass
and shaping the CSM structure.

Our simulations produce output for the ejecta at each timestep. The top panel of \Figref{fig:sptime} shows the cumulative mass lost to PPI-driven
pulses for our example $M_\mathrm{He,init}=50\,M_\odot$ He core. In this specific model, the total
(H-free) ejecta mass is about $1.2\,M_\odot$ by the end of the evolution. Had our star retained an H-rich envelope until
the onset of the first pulse, the remaining H-rich envelope at the onset of the
instability would likely add to the amount of mass in the CSM (see also Appendix~\ref{sec:hrichcomp}).

The bottom panel of \Figref{fig:sptime} shows the density distribution around
the star as a function of distance from the star (y-axis) and time until the final CC
(x-axis). The cyan line shows the radius of the bound material
(cf.~the red curve in \Figref{fig:R_t}), which we assume to be the initial radius from
which the ejecta are launched. To compute the CSM density we assume propagation of the ejecta at constant velocity. We use the
velocity at the time the material first exceeds the local escape velocity 
as computed by \texttt{MESA}, and it is
typically a few thousand $\kms$. We return
to the ejecta velocity in \Secref{sec:grid_ejecta}.
Assuming a constant velocity for the propagation of the ejecta corresponds to
neglecting radiative cooling, internal collisions of the ejecta, and
multi-dimensional effects \citep[e.g.,][]{chen:19}, and we discuss it here only for
illustration purposes. Our output files\footnote{Publicly available at \href{https://doi.org/10.5281/zenodo.3406356}{https://doi.org/10.5281/zenodo.3406356}.} contain the amount of mass ejected, its
center-of-mass velocity, chemical composition and thermal state (averaged
by mass over all the mesh points that exceed the local escape
velocity in the current timestep), which could be used as input for more
detailed simulations to predict the CSM structure around
PPI+CC models. This ejecta output neglects the possibility of fallback, which
however could be implemented when using these files as input for hydrodynamical
simulations of the CSM.

The CSM structure shown in \Figref{fig:sptime} for our example model
shows H-poor/He-rich CSM starting from $\sim$\,$10^{13}\,\mathrm{cm}$
and extending out to $\sim$\,$10^{14}\,\mathrm{cm}$. The CSM densities
reach $10^{-5}-10^{-4}\,\mathrm{g\ cm^{-3}}$. These value are typical
for the models in our grid, and fall in the range of CSM distances and
densities inferred from transient observations
\citep[e.g.,][]{gomez:19}.

However, mass ejections that happen in subsequent timesteps (possibly
with no mass ejected in between) in \texttt{MESA} might not be
physically distinct events. To count the mass-ejection events, we need
to group mass ejections in timesteps separated by less than a
dynamical timescale in one individual event. Moreover, one mass
ejection event can last several dynamical timescales, for example the
final mass ejection and full disruption of a PISN is expected to
produce a long transient \citep[e.g.,][]{gal-yam:09}. We estimate the
dynamical timescale as the free-fall timescale
$\tau_\mathrm{ff}=2\pi\sqrt{R_\mathrm{photo}^3/GM_\mathrm{bound}}$
where G is Newton's constant, and $R_\mathrm{photo}$ and
$M_\mathrm{bound}$ are the (time-dependent) photospheric radius and
mass gravitationally bound to the star. We define the beginning of a
mass loss event as the timestep during which at least
$10^{-6}\,M_\odot$ has been removed from the star since the last mass
loss event (either in one timestep, or cumulatively). We require each
mass ejection event to last at least one free fall timescale
(calculated at the beginning of the pulse), and define its end as soon
as the amount of mass to be ejected in the following 100
$\tau_\mathrm{ff}$ (now calculated at the end of the pulse) is less than
$10^{-7}\,M_\odot$. This last condition allows us to count as a single
event mass ejection episodes that last longer than a dynamical
timescale. All together, these requirements enforce that ejections
which numerically happen in different timesteps separated by less than
a free fall timescale are not counted as separate events.

Adopting
this criterion to count the mass ejection events, our
$M_\mathrm{He,init}=50\,M_\odot$ model only has one mass-ejection
episode (cf. tens of core ignitions, see \Secref{sec:thermo_def}),
which starts roughly
$\sim\,0.015\,\mathrm{years}\simeq130\,\mathrm{hours}$ before CC.

We can tentatively apply the same threshold to define the
  beginning of an explosion for a terminal PISN with full
  disruption. In this case, the
duration of the PISN events in our grid exceeds months even for the
least massive PISN model with $M_\mathrm{He,init}=80.75\,M_\odot$ in
agreement with previous studies. Our stopping conditions do
not allow models to reach what would appear as the observational end of
a PISN.

\section{Pulsational pair-instability-generated CSM}
\label{sec:grid_ejecta}

\begin{figure}[bp]
  \centering
  \includegraphics[width=0.5\textwidth]{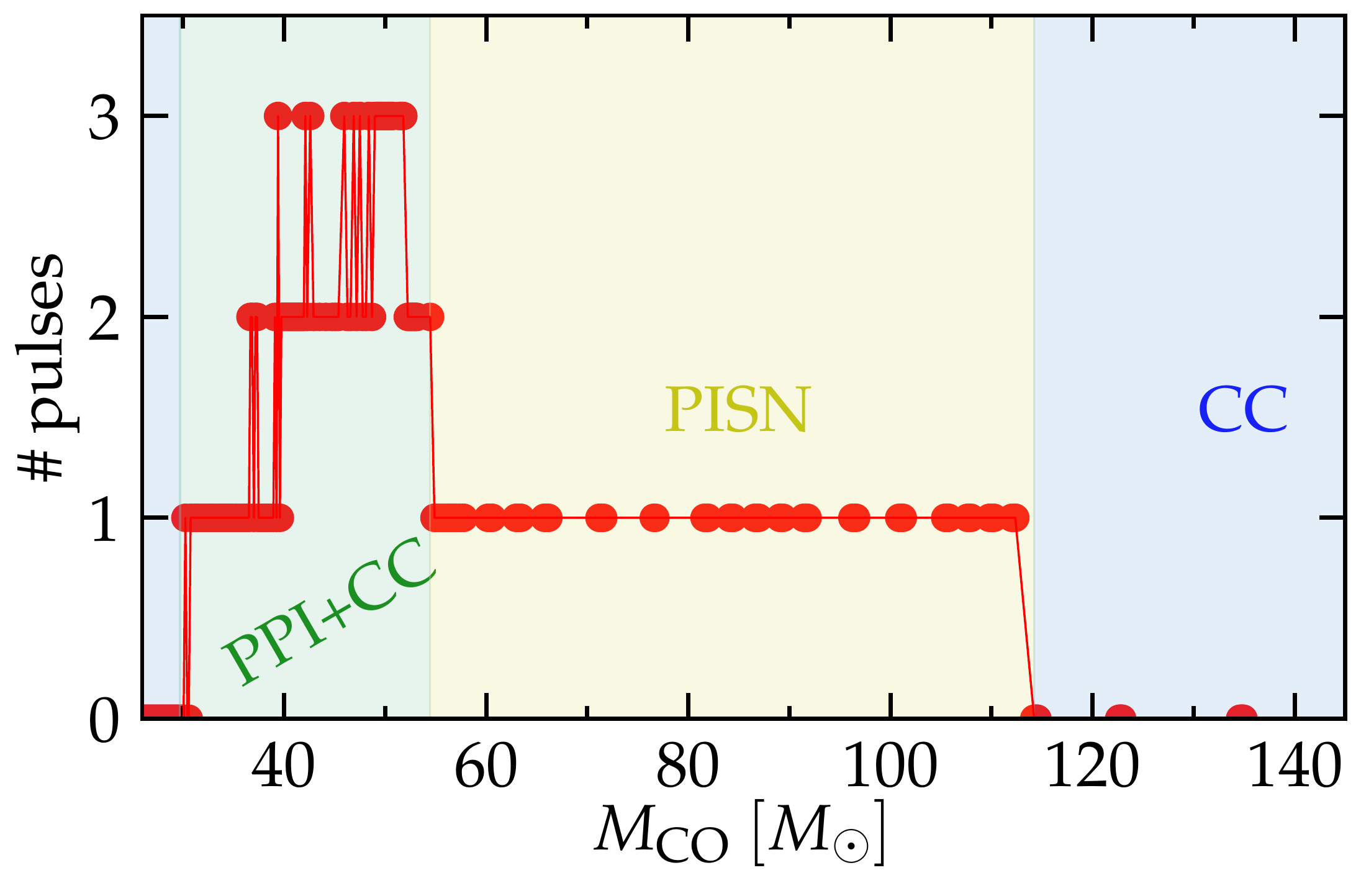}  
  \caption{Number of mass-ejection events caused by pair instability as a
    function of CO core mass. The color shading indicates the
    approximate range for each behavior: core collapse without experiencing
    PPI-driven mass loss (CC, blue), PPI-driven mass loss (PPI+CC, green), or
    full disruption in a PISN (yellow), which we define as one mass loss event.
    The noisiness is caused by the occurrence of mass loss event right at the
    time of the final core collapse.}
  \label{fig:num_pulses}
\end{figure}

\begin{figure}[tp]
  \centering
  \includegraphics[width=0.5\textwidth]{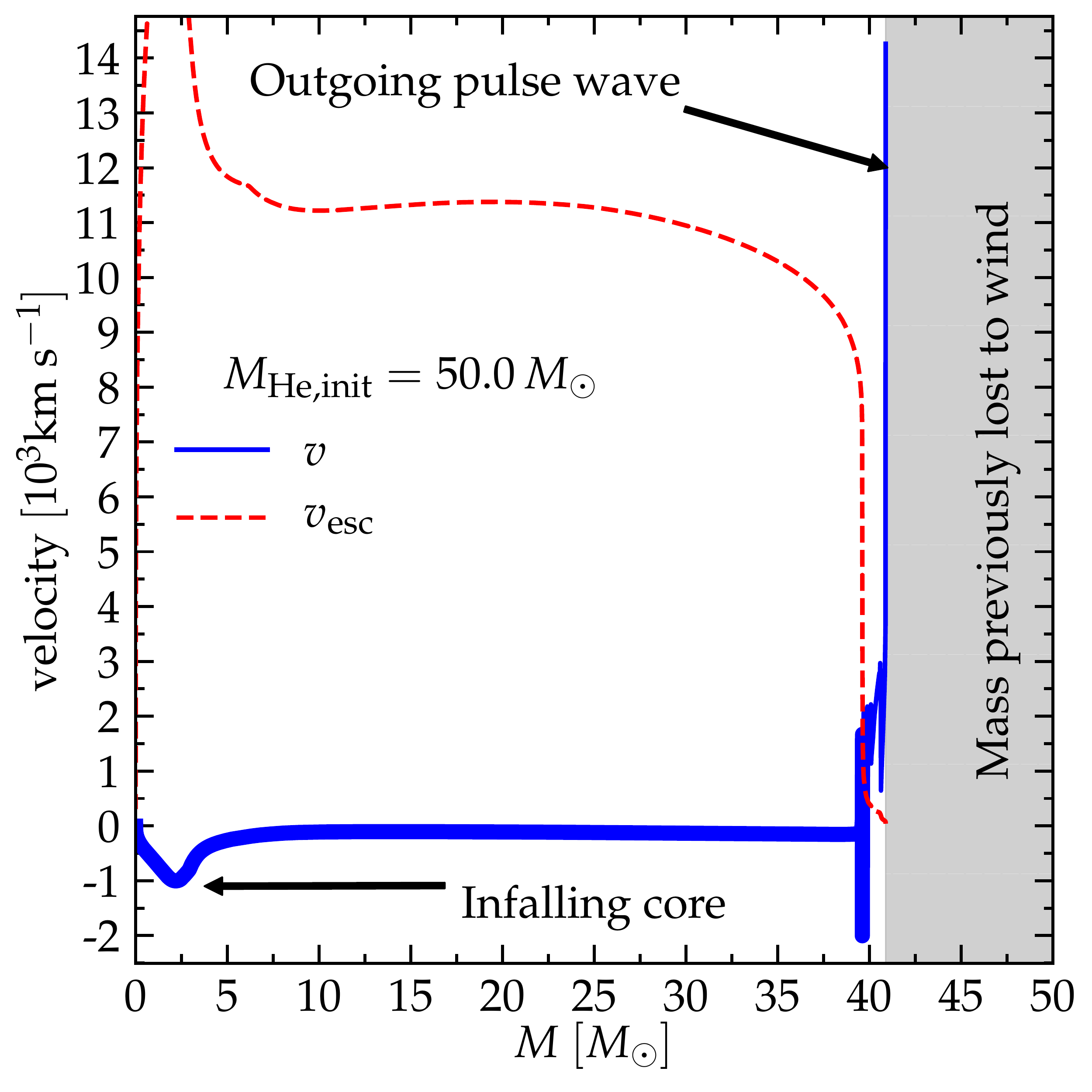}
  \caption{Velocity profile at the onset of core-collapse for a 50\,$M_\odot$ He
    core, which undergoes a PPI mass-loss event while collapsing. The dashed red
    line shows the escape velocity profile, the thick blue
    line indicates the profile of the bound material, while the thinner line
    represents material still on the Lagrangian mass grid, but already beyond
    the escape velocity. The gray area indicates the model dependent
    amount of mass lost to stellar winds.}
  \label{fig:velocity_profile}
\end{figure}

We now discus the CSM that can be created by PPI evolution across
our grid of models. In principle, this CSM can be probed by time-domain
observations, if the final core-collapse causes an associated explosion
(see~\Secref{sec:explodability}), or because of collisions
  between shells ejected at different times
  \citep[e.g.,][]{woosley:07,woosley:17, woosley:19}. Most of
the CSM properties do not depend on which criterion is used to
define the beginning or end of the pulses, except the number of pulses and their
duration. For these quantities, we adopt the definition of
\Secref{sec:mejection_def} based on the ejection of matter, which is the most
relevant for discussing the CSM structure.

\Tabref{tab:ejecta} summarizes the time, duration, and amount of mass loss in
each event for all the PPI+CC models in our grid, and \Figref{fig:num_pulses}
shows the number of pulses as mass-ejection events contributing to the CSM.
Models evolving to CC without any mass ejection have zero pulses. We
define full disruption in a PISN as a one-pulse event, although these would
not contribute to the CSM itself. The color in the background emphasizes
the various evolutionary behaviors, but using the definition from
\Secref{sec:mejection_def} the mass threshold separating CC (blue)
from PPI+CC (green) evolution is well defined at
$M_\mathrm{He,init}=40.5\,M_\odot$, correponding to $M_\mathrm{CO}=30\,M_\odot$.

\begin{figure}[htbp]
  \centering
  \includegraphics[width=0.5\textwidth]{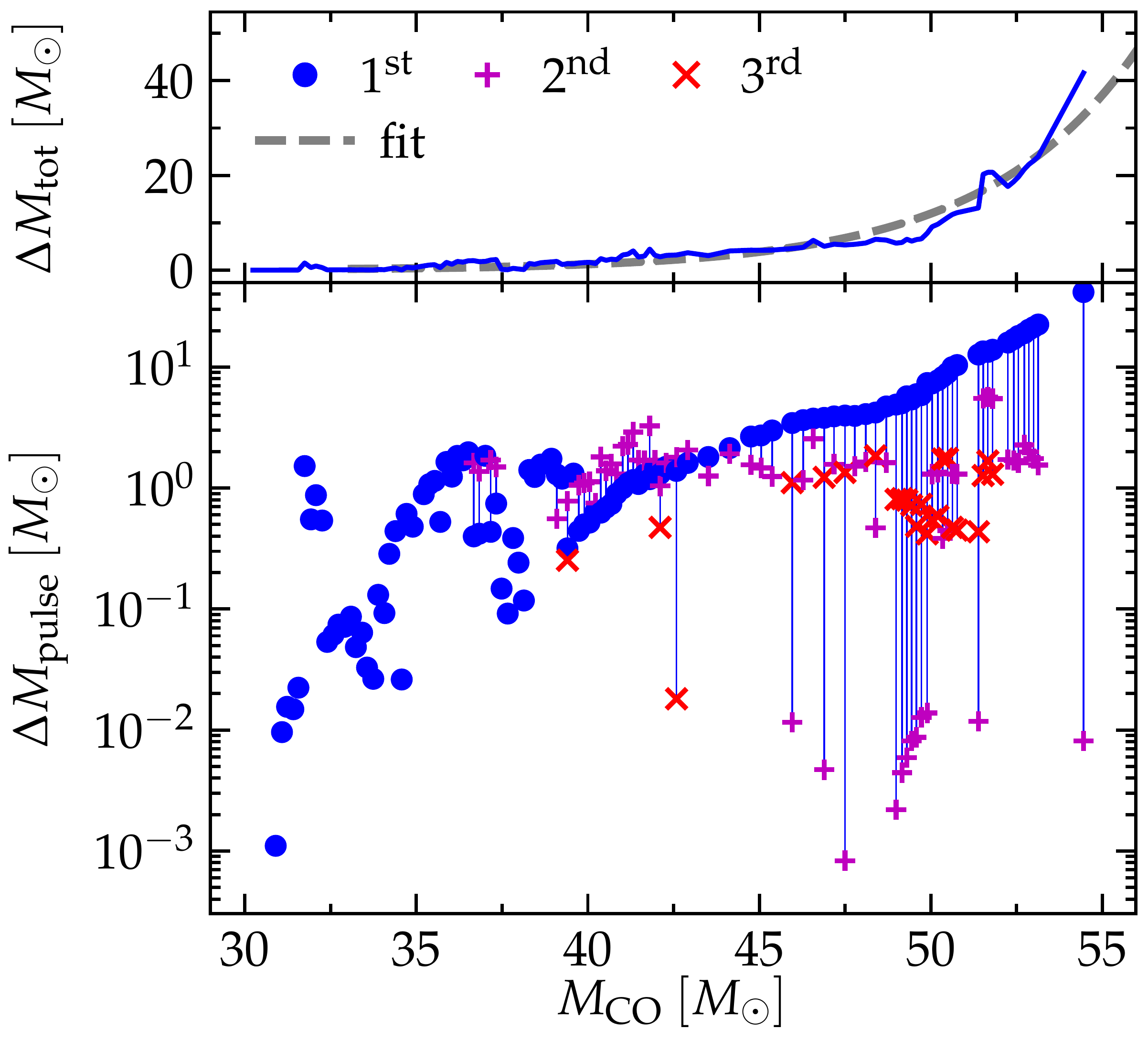}  
  \caption{PPI-driven mass loss as a function of the CO core mass. The top panel shows the total mass ejected in
    pulses, the bottom panel shows the mass lost in individual pulses. The
    amount of mass lost does not have a monotonic behavior with pulse number,
    and spans a wide range of values. The first
    pulse is shown as a blue dot, and the second and third, if they
    occur, are shown as a
    purple plus and red cross, respectively. Thin vertical lines connect multiple pulses for the same $M_\mathrm{CO}$.}
  \label{fig:dm_pulses}
\end{figure}

The number of pulses is zero at the lower end, and increases up to three
distinct mass-ejection events for the central part of the PPI+CC mass range. At
even higher masses, approaching the PPI+CC/PISN boundary, the number of pulses
decreases again, although the amount of mass ejected increases (see also
\Figref{fig:dm_pulses}): this is because pulses become more energetic and
consume more nuclear fuel at once \citep[e.g.,][]{woosley:07, chatzopoulos:12b,
  woosley:17, woosley:19}.

The green region shows some noise in the number of pulses: the reason for this
is illustrated in \Figref{fig:velocity_profile}, which shows the velocity as a function
of Lagrangian mass coordinate for our $M_\mathrm{He,init}=50\,M_\odot$
core at the onset of core collapse. Many models exhibit a similar behavior, with
an outgoing pulse wave at the onset of CC: this means that the PPI mass
ejection is still going on while the Fe core starts collapsing.

\Figref{fig:dm_pulses} summarizes the amount of mass lost to PPI-driven pulses
across our model grid. The bottom panel shows the amount of mass lost per
individual pulse, the pulse number is represented by the color of the
filled circles.

The typical amount of mass lost varies from
$\lesssim 10^{-3}\,M_\odot$ for the lowest-mass models ejecting some
mass, up to $\simeq20\,M_\odot$ (of He-rich material) at the upper
mass end, just below the minimum mass for PISN. For models producing
more than one mass ejection event (i.e., the models for which also a
purple and possibly a red dot are shown), the amount of mass lost per
pulse does not behave monotonically with the pulse number. For
$50\,M_\odot\lesssim M_\mathrm{He,init}\lesssim 62\,M_\odot$,
corresponding to
$36\,M_\odot\lesssim M_\mathrm{CO}\lesssim 43\,M_\odot$ the second
pulse (purple) ejects more mass than the first (blue), while for
higher masses the first pulse removes more mass than the second.

\begin{figure}[htbp]
  \centering
  \includegraphics[width=0.5\textwidth]{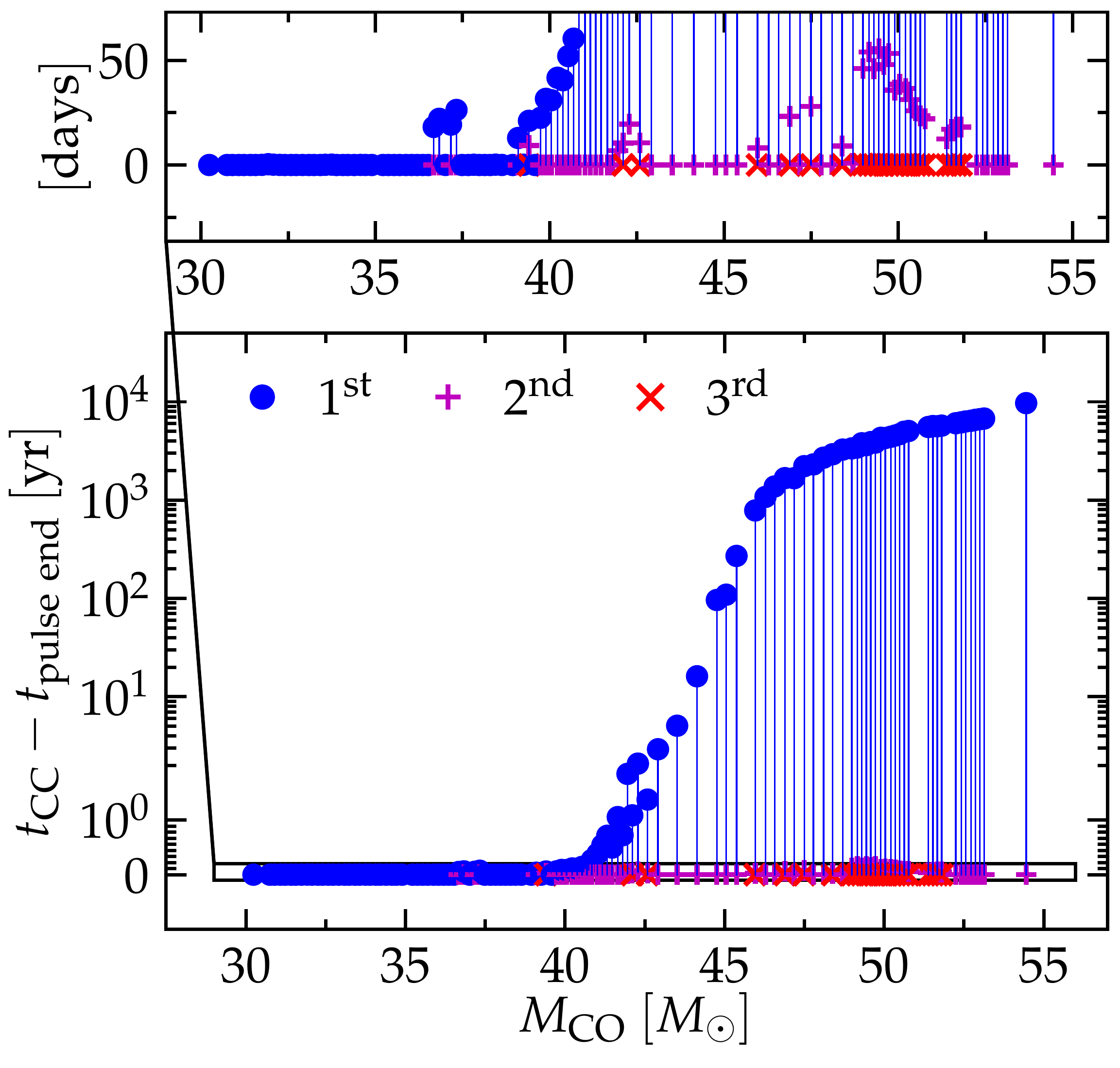}
  \caption{Delay time between the final core-collapse and the end of
    each mass loss event in a PPI pulse. Typical delays are on the order of few months,
 but they increase steeply with the initial He core mass of the PPI
 progenitor, which produce fewer but more energetic pulses, up to
 $10^4$\,years. The top inset magnifies the range of half a year
 delays on a days scale. The first pulse is shown as a blue dot, and
 the second and third, if occurring, are shown as a purple plus and
 red cross, respectively, with a thin blue line connecting the pulses
 of the same model.
}\label{fig:timing}
\end{figure}

The top panel of \Figref{fig:dm_pulses} shows the total amount of mass lost to
PPI ejecta, i.e., the sum of the mass ejected in each individual pulse. A trend
of more massive models producing more energetic pulses and driving more mass
loss is evident. We provide a simple
fitting formula for the total amount of He-rich mass lost in PPI-driven events for
$33\,M_\odot\leq M_\mathrm{CO}\leq56.5\,M_\odot$ which produce a CSM mass larger than $\sim$\,$0.2\,M_\odot$, shown as a dashed gray line in the top panel of
\Figref{fig:dm_pulses}:

\begin{equation}
  \label{eq:fit_dmtot}
  \frac{\Delta M_\mathrm{tot}}{M_\odot} = 0.000147\times 10^{0.098 M_\mathrm{CO}/M_\odot} \ \ \mathrm{for} \ \ 33\,M_\odot \leq M_\mathrm{CO}\leq 56.5\,M_\odot \ \ .
\end{equation}

The total
mass lost to pulses should be added to the amount of mass lost due to winds to
calculate the total mass in the CSM. The density distribution of the
CSM generated by PPI-driven pulses and wind mass loss are likely to be very
different from each other. In cases where stars retain a (loosely bound) H-rich envelope at the onset
of the first pulse, the mass of such an envelope at the start of the pulses should also be added to the total
mass lost in the first mass loss event.

\Figref{fig:timing} shows the delay time between the end times mass ejections and the final CC, as a
function of the CO core mass of our models. The timing of the mass-loss events also spans a large range, from
zero (see also \Figref{fig:velocity_profile} for example) to $\sim$\,$10^{4}$\,years,
corresponding roughly to the Kelvin-Helmholtz timescale of the most massive
PPI+CC. We emphasize that many models in our grid show submonth delays between the
last pulse and the final CC, which makes them
candidates for the detection of CSM interactions in early observations
of the SN explosions (see also inset in \Figref{fig:timing}).

The delay (in between pulses and between each pulse and CC) increases with the
core mass, because more massive models produce more energetic pulses that
drive the star farther from gravo-thermal equilibrium, increasing the amount of
time needed to return to equilibrium after a pulse and resume the
final evolution. While typically for very massive stars the neutrino luminosity
greatly exceeds their photon luminosity $L_\nu\gg L$ \citep[they are ``neutrino
stars'',][]{fraley:68}, this is not always true for the most massive PPI+CC
models. For these, the adiabatic expansion of the core can
leads to central temperature and densities too low for significant neutrino
cooling to occur. Thus, after a pulse begins, these models
transition from evolving on a neutrino-mediated thermal timescale ($\propto GM^2/RL_\nu $)
to a photon-mediated thermal timescale ($\propto GM^2/RL$) in between pulses, which increases their
interpulse time.

\begin{figure}[htbp]
\centering
  \includegraphics[width=0.5\textwidth]{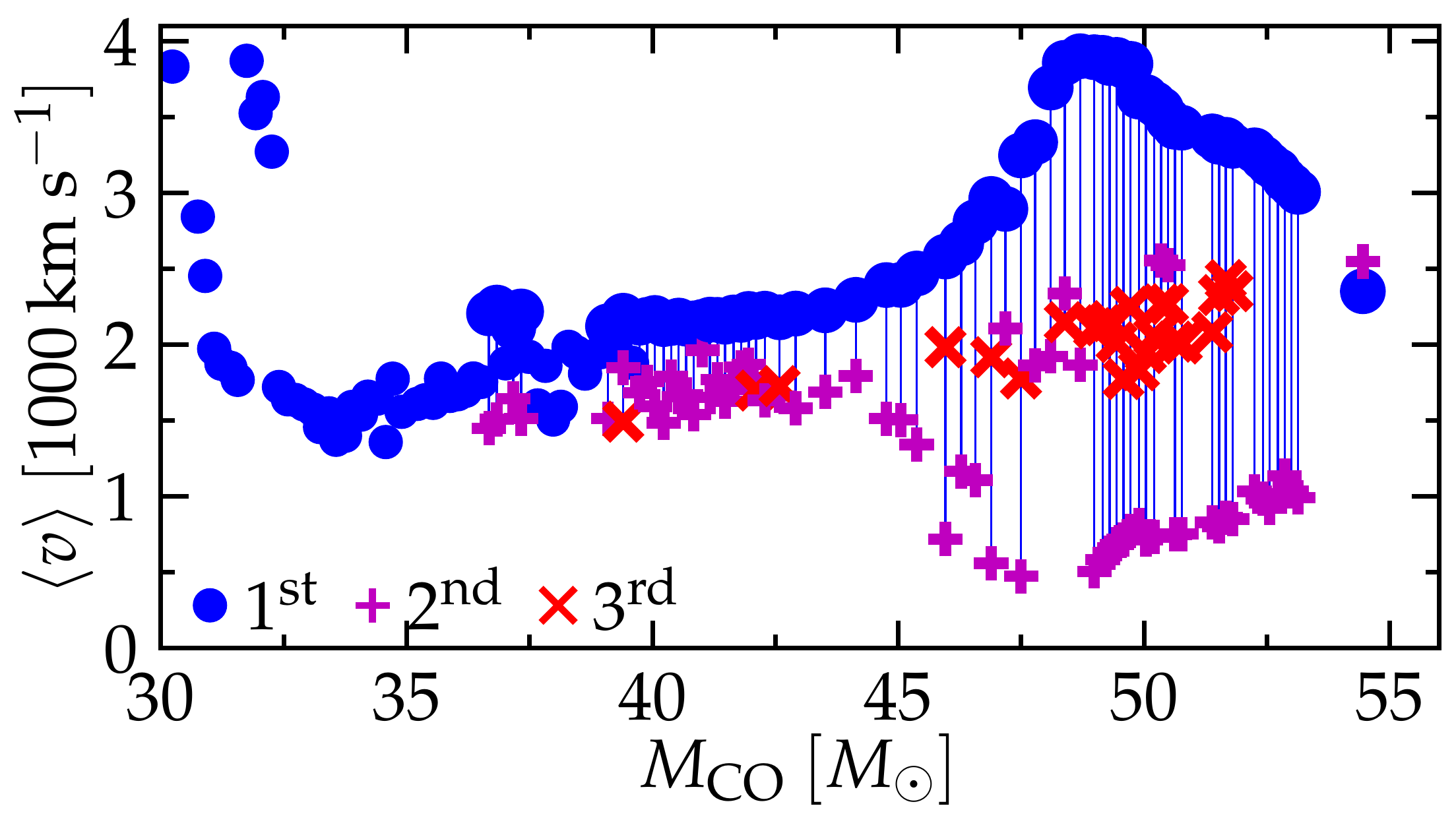}
  \caption{Center-of-mass velocity of the layers ejected at each pulse. The
    first pulse (blue filled circles) produces typically higher ejection velocities, and the
    ejecta move at $\sim$\,few thousand $\kms$ which suggests a connection with
    (some subclasses) of SN~Ibn.}
  \label{fig:v_ejecta}
\end{figure}

Figure~\ref{fig:v_ejecta} shows the center-of-mass velocity of the layers
ejected (which is calculated as the mass-weighted average of the center-of-mass velocity of
the layers ejected at each timestep over the duration of the mass ejection
event). Unlike the other quantities characterizing a pulse, the ejecta
velocities we find do not span orders of magnitude, and are typically a few
$\sim$\,$1000\,\kms$. This suggests that mass ejection during a PPI might explain
the He-rich circumstellar material required to explain at least
some of the spectra of SN~Ibn showing narrow He emission lines
\citep[e.g.,][]{pastorello:08}, provided that there is a way to
excite the ejected shells. This could be due to collisions
between the third and second pulse within the first ejected
shell, or because of a successful explosion at the final CC (see also \Secref{sec:explodability}).

The first pulse (blue filled circles) are almost always faster than the
later pulses. Conversely, when they occur, the third pulse (red crosses) is often faster
than the second (purple pluses). Therefore, many models might result in collisions in between the ejecta which can
appear as SN impostors, as noted by \cite{woosley:07} and \cite{woosley:17}.

We only report the center-of-mass velocity for the ejected
shells, because our treatment of the ejecta is presently very
simplified. Radiation-hydrodynamics simulations of the ejecta
propagation would be desirable to properly quantify the velocity
distribution of the ejecta, and in particular quantify the
low-velocity tail which will appear first in observations of CSM interactions.

\section{Explodability of CC and PPI+CC models}
\label{sec:explodability}

Does the final CC of a post-PPI star result in a successful explosion?
This question remains relevant because a terminal
  explosion would potentially allow us to probe all the layers ejected
  previously. Whether the final collapse of PPI+CC evolution is
  accompanied by a successful, albeit possibly weak, explosion
likely constitutes the biggest uncertainty underlying this study,
and it is not a question that we can settle using only our stellar
structure models.
We emphasize that even in the absence of a terminal explosion,
electromagnetic transients from PPI are possible and expected
because of collisions between shells \citep[e.g.,][and
Sec.~\ref{sec:grid_ejecta}]{woosley:07, woosley:17, woosley:19}.

Based on the extrapolation of one-dimensional parametric CC
simulations, the typical expectation is that He cores with masses
larger than $\gtrsim 10\,M_\odot$ fail to explode
\citep[e.g.,][]{oconnor:11, ugliano:12, ertl:16}, therefore a
successful terminal explosion in PPI+CC models appears
unlikely. However, the observations of (BH) X-ray binary kinematics
\citep[e.g.,][and references therein]{brandt:95b, fragos:09, atri:19},
and possible spin misalignment in gravitational wave events
\citep[e.g.,][]{oshaughnessy:17} might require BH natal kicks. These
can occur only with an explosion and some level of asymmetry in the
ejecta \citep[e.g.,][or possibly in the neutrino flux]{janka:13,
  janka:17, chan:18, chan:20}. Thus, the current understanding of
BH formation cannot yet be considered final.

As far as we are aware, there are no published multi-dimensional
calculations investigating the final “explodability” of PPI+CC
structures. Most of the available studies \citep[e.g.,][]{ott:18,
  kuroda:18} target much lower mass progenitors (up to
$\sim$$70\,M_\odot$ including the H-rich envelope) and/or
rely on artificially imposed large perturbations or enhanced
neutrino-nucleon scattering to obtain explosions \citep[e.g.,][]{chan:18, chan:20}.
Therefore, these results may not meaningfully extrapolate to the PPI+CC regime.

Even in the case of successful but weak explosions, these studies either do not follow the explosion long enough
to study the amount of mass ejected (if any), or find that only a fraction of the envelope is ejected \citep[e.g.,][]{chan:20}. While our models do not have any H-rich envelope to begin with, the He core at precollapse can reach radii similar to that of a H-envelope in yellow-supergiant stars.  Specifically, the radii of bound material
($v<v_\mathrm{esc}$) at the onset of CC range between a few to a few hundreds of solar radii
($\sim10^{11}-10^{13}\,\mathrm{cm}$). Such extended He cores, effectively akin to He envelopes in some cases, have significantly different density profiles compared to the stellar structures explored in the aforementioned studies.  It is unclear whether the usual definition of the core-envelope boundary, based on abundances, makes sense from a core-collapse dynamics perspective for our models.

Although also unexplored and uncertain in this regime, even very weak
explosions due to the initial neutrino emission before the formation
of an event horizon (e.g.,
\citealt{nadezhin:80,lovegrove:13,coughlin:18} and possibly observed
by \citealt{adams:17}) might potentially be sufficient to start an
electromagnetic transient when the outermost layers hit the previously
ejected mass. Once again, this mechanism is thought to only
  eject loosely-bound H-rich outer layers at low velocity, and in red
  supergiant progenitors of much lower mass than we explore here. The
  large radial extension of some of our precollapse models might
  possibly result in the ejection of some equally loosely bound
  material, but the typical amount of mass with binding energy
  sufficiently low is only of order $0.01\,M_\odot$. Should this
  scenario produce some ejecta, the amount of kinetic energy
available seems unlikely to produce transients detectable
  at large distances \citep{chevalier:12}.

Explosion mechanisms relying on rotational energy
\citep[e.g.,][]{moesta:15} and jets \citep[e.g.,][]{gilkis:16,
  soker:19}, which have also been proposed to explain type Ic SNe
showing broad lines \citep[e.g.,][]{barnes:18} might also deserve
attention in this context. In \cite{marchant:19} we showed that
PPI-driven mass loss does not dramatically decrease the core angular
momentum. However, we are not aware of any ``explodability'' criterion
for this kind of core-collapse engine.

One key difference between PPI+CC cores and lower mass stars
contributing to the differences in density profiles is the
production of $^{56}\mathrm{Ni}$ during pulses, before the final CC.
We emphasize
that the energy released by $^{56}\mathrm{Ni}$ decay is not expected to play a
role \emph{during} core-collapse, as the half-life of
$^{56}\mathrm{Ni}$ is much longer than the few hundred millisecond
timescale of CC. Its effect is to modify the initial
conditions for CC during the interpulse evolution of the models.

The typical outermost mass coordinate
where we find a mass fraction of $^{56}\mathrm{Ni}$ larger than 0.01
is less than $5\,M_\odot$, thus we expect most of the
$^{56}\mathrm{Ni}$ will eventually be accreted into the final
BH. However, the $^{56}\mathrm{Ni}$ is produced weeks, months, or up
to several years before CC. This provides sufficient time for the
$^{56}\mathrm{Ni}$ to decay and causes the core to expand due to this
heating. This makes the precollapse core structure of PPI+CC models
qualitatively different from lower mass models routinely used for CC
simulations. Detailed simulations of the final CC structure
(including large nuclear networks to capture the precollapse
deleptonization, \citealt{farmer:16,renzo:17}) and explosion are
needed to shed light on what the these qualitative differences
mean for the ``explodability'' of PPI+CC models.  In the most
massive PPI+CC models, the decay of the $^{56}\mathrm{Ni}$ produced
during pulses can ultimately unbind what is left of the core
resulting in a minimum BH mass that can be obtained via PPI+CC
\citep[see also][]{marchant:19}.

\begin{figure}[tbp]
  \centering
  \includegraphics[width=0.5\textwidth]{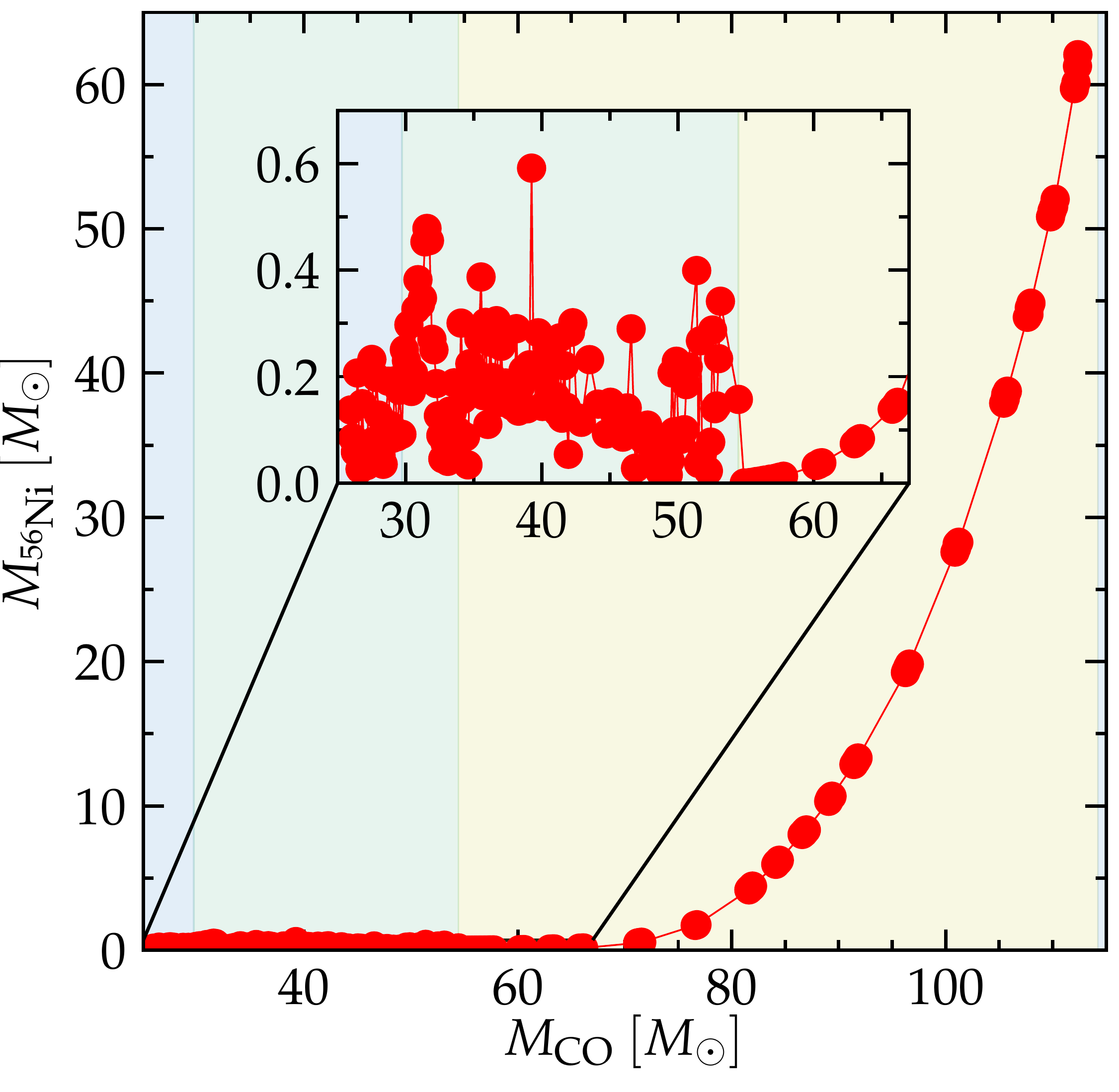}
  \caption{$^{56}\mathrm{Ni}$  mass present in the deep interior of
      the star at the onset of CC or
    PISN. The color background is the same as \Figref{fig:num_pulses} and
    indicates the approximate evolution (CC for blue, PPI+CC for green, and PISN
    for yellow). We computed these models using a 22-isotope nuclear reaction
    network.}
  \label{fig:ni}
\end{figure}

\Figref{fig:ni} shows the amount of $^{56}\mathrm{Ni}$ present in our
PPI+CC and PISN models at the end of the evolution: for PPI+CC models
the total mass of $^{56}\mathrm{Ni}$ is typically
$M_\mathrm{Ni}\simeq0.2-0.4\,M_\odot$, i.e., about one order of
magnitude more than what is produced in typical core-collapse SNe
\citep[e.g.,][]{wongwathanarat:13}. This value increases steeply in
the PISN range, reaching about $\sim$\,$60\,M_\odot$ at the upper end,
in good agreement with \cite{heger:02} and \cite{woosley:02}
results. However, our calculations are based on a
22-isotope nuclear reaction network which is known to produce
$M_\mathrm{Ni}$ deviating by up to a factor of $\sim$\,1.5$\times$ in
either direction from results computed with larger nuclear reaction
networks (regardless of the final fate of the models between CC,
PPI+CC, or PISN).

\begin{table}[!htbp]
  \centering
  \caption{He core mass at the beginning of our simulations
      ($M_\mathrm{He,init}$) and at He depletion
      ($M_\mathrm{He,depl}$), carbon-oxygen core mass ($M_\mathrm{CO}$), iron core mass ($M_\mathrm{Fe}$), mass location
    $M_4$ where the specific entropy is lower than
    $4k_BN_A$, and mass gradient $\mu_4\equiv dM/dr|_{s=4k_BN_A}$ for a representative
    subset of our models eventually forming a BH at
    $Z=0.001$. All the
    models we computed are available at \href{https://doi.org/10.5281/zenodo.3406356}{https://doi.org/10.5281/zenodo.3406356}.  \label{tab:expl}}
  \begin{tabular}[htbp]{l|c|ccccc}
    & $M_\mathrm{He,init}$ &$M_\mathrm{He,depl}$ & $M_\mathrm{CO}$ &
                                                                     $M_\mathrm{Fe}$ & $M_4$ & \multirow{2}{*}{$\mu_4$}\\
    & $[M_\odot]$& $[M_\odot]$& $[M_\odot]$& $[M_\odot]$& $[M_\odot]$& \\\hline\hline
      \multirow{7}{*}{
  \begin{sideways}
    CC    
  \end{sideways}}
 &  35.00 &  30.03  &  25.96 &  1.82  &  2.22  &  0.16\\
 &  36.00 &  30.78  &  26.64 &  1.92  &  2.30  &  0.18\\
 &  37.00 &  31.53  &  27.33 &  1.75  &  2.33  &  0.18\\
 &  38.00 &  32.27  &  28.00 &  1.59  &  2.34  &  0.19\\
 &  39.00 &  33.02  &  28.71 &  1.65  &  2.39  &  0.21\\
 &  40.00 &  33.75  &  29.38 &  2.07  &  2.44  &  0.25\\
 &  41.00 &  34.49  &  30.05 &  1.96  &  2.48  &  0.28\\\hline
    \multirow{41}{*}{
    \begin{sideways}
      PPI+CC    
    \end{sideways}}
    
 &  42.00 &  35.22  &  30.75 &  1.60  &  2.53  &  0.32\\
 &  43.00 &  35.95  &  31.41 &  1.67  &  2.57  &  0.36\\
 &  44.00 &  36.68  &  32.07 &  2.17  &  2.62  &  0.11\\
 &  45.00 &  37.40  &  32.73 &  2.18  &  2.68  &  0.26\\
 &  46.00 &  38.12  &  33.42 &  1.76  &  2.72  &  0.19\\
 &  47.00 &  38.83  &  34.07 &  1.92  &  2.77  &  0.20\\
 &  48.00 &  39.55  &  34.72 &  1.92  &  2.82  &  0.28\\
 &  48.25 &  39.73  &  34.90 &  2.17  &  2.82  &  0.23\\
 &  48.75 &  40.08  &  35.22 &  1.72  &  2.86  &  0.27\\
 &  49.00 &  40.27  &  35.38 &  1.56  &  2.86  &  0.33\\
 &  50.00 &  40.98  &  36.04 &  2.33  &  2.90  &  0.28\\
 &  51.00 &  41.69  &  36.67 &  1.69  &  2.64  &  0.56\\
 &  52.00 &  42.38  &  37.33 &  2.09  &  2.77  &  0.39\\
 &  53.00 &  43.09  &  37.98 &  1.96  &  2.57  &  0.21\\
 &  54.00 &  43.79  &  38.63 &  1.95  &  2.56  &  0.24\\
 &  54.50 &  44.14  &  38.94 &  1.97  &  2.54  &  0.21\\
 &  54.75 &  44.31  &  39.09 &  1.84  &  2.51  &  0.29\\
 &  55.00 &  44.49  &  39.25 &  1.80  &  2.42  &  0.30\\
 &  56.00 &  45.19  &  39.89 &  2.20  &  2.25  &  0.52\\
 &  57.00 &  45.88  &  40.53 &  2.18  &  1.90  &  0.45\\
 &  58.00 &  46.57  &  41.17 &  2.21  &  2.56  &  0.45\\
 &  59.00 &  47.26  &  41.80 &  2.25  &  2.26  &  0.49\\
 &  60.25 &  48.13  &  42.59 &  2.18  &  2.31  &  0.48\\
 &  61.75 &  49.14  &  43.51 &  1.65  &  2.24  &  0.29\\
 &  62.75 &  49.80  &  44.14 &  1.99  &  2.25  &  0.42\\
 &  63.75 &  50.48  &  44.76 &  2.03  &  2.28  &  0.32\\
 &  64.25 &  50.81  &  45.05 &  1.68  &  2.24  &  0.41\\
 &  65.75 &  51.81  &  45.96 &  2.04  &  2.25  &  0.33\\
 &  66.25 &  52.15  &  46.28 &  1.63  &  2.24  &  0.39\\
 &  67.25 &  52.80  &  46.89 &  1.94  &  2.25  &  0.29\\
 &  68.25 &  53.46  &  47.50 &  2.05  &  2.18  &  0.27\\
 &  69.25 &  54.12  &  48.10 &  2.04  &  2.21  &  0.31\\
 &  70.25 &  54.78  &  48.70 &  2.02  &  2.17  &  0.27\\
 &  71.00 &  55.27  &  49.16 &  2.03  &  2.25  &  0.32\\
 &  72.00 &  55.91  &  49.72 &  2.31  &  2.16  &  0.23\\
 &  73.00 &  56.56  &  50.35 &  2.16  &  2.19  &  0.42\\
 &  74.75 &  57.68  &  51.39 &  1.76  &  2.02  &  0.73\\
 &  75.00 &  57.85  &  51.52 &  2.02  &  2.12  &  0.25\\
 &  76.25 &  58.64  &  52.24 &  2.15  &  2.15  &  0.39\\
 &  77.00 &  59.13  &  52.72 &  2.19  &  2.15  &  0.52\\
 &  80.00 &  61.03  &  54.45 &  2.09  &  2.31  &  0.53\\
    \hline
    \multicolumn{7}{c}{PISN}\\\hline
    \multirow{3}{*}{
  \begin{sideways}
    CC    
  \end{sideways}}
 &  200.00 &  125.26  &  114.23 &  31.50  &  --  &  --\\
 &  220.00 &  134.22  &  122.64 &  10.78  &  --  &  --\\
 &  250.00 &  147.03  &  134.60 &  5.77  &   --  &  --\\
  \end{tabular}
\end{table}

\Tabref{tab:expl} lists, for a representative subset of models,
quantities commonly used to determine the ``explodability'' of a
stellar model. These focus on the innermost layers of the star,
  but already include the effect of the precollapse decay of
  $^{56}\mathrm{Ni}$ produced during pulses in this region of the star. Specifically, we report the initial He core mass
$M_\mathrm{He,init}$ and its value at the end of He core
  burning\footnote{Since our models do not have a H-rich
    envelope, this is the total mass when the central He mass fraction
    reaches $X_c(^4\mathrm{He})<10^{-5}$. We emphasize that this value is sensitive
    to the adopted wind mass loss rates.}
$M_\mathrm{He, depl}$, the CO core mass $M_\mathrm{CO}$, the
final iron core mass $M_\mathrm{Fe}$, defined as the outermost location where the mass
fraction of $^{28}\mathrm{Si}\leq0.01$ and the mass fraction of
Fe-group elements, i.e., with more than 46 nucleons, exceeds 0.1; and
the two
parameters 
proposed by \cite{ertl:16}. These are the mass coordinate $M_4$ at
which the specific entropy reaches $4k_BN_A$, where $k_B$ is
Boltzmann's constant and $N_A$ is Avogadro's number, and the mass
gradient at this location $\mu_4$ \citep[Eq.~6 in][]{ertl:16}. Since
the nuclear reaction network we employ does not allow for detailed
treatment of the electron captures and $\beta$ decays, which determine
the final electron-to-baryon ratio, we avoid listing the compactness
parameter \citep[e.g.,][]{oconnor:11}, which is sensitive to these
modeling assumptions \citep[][]{farmer:16,renzo:17}. We caution that
three-dimensional simulations \citep[e.g.,][]{ott:18,kuroda:18,chan:18,chan:20} might
give a different outcome than 1D parametric simulations used to assess
the explodability of grids of models
\citep[e.g.,][]{oconnor:11,ugliano:12,mueller:19,couch:19}.

\begin{figure}[tbp]
  \centering
  \includegraphics[width=0.5\textwidth]{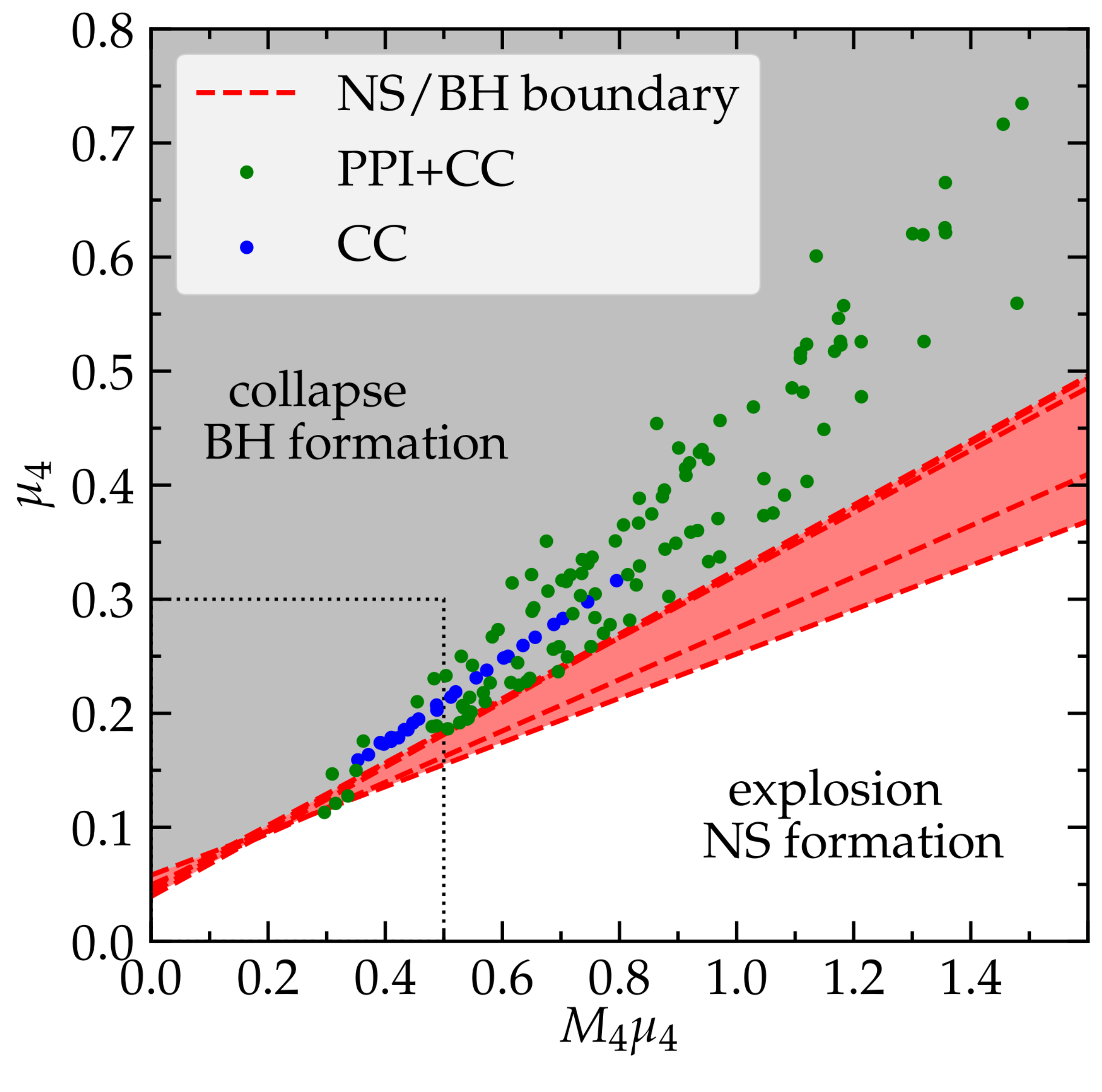}
  \caption{CC and PPI+CC models on the Ertl criterion plane for
    "explodability". The red area indicates the uncertain boundary
    region between successful explosions and neutron star formation
    (white) and collapse with BH formation (gray). The dotted
    rectangle in the bottom left indicates the range originally showed in \cite{ertl:16},
    most of our models require extrapolating outside this range. Blue
    dots correspond to CC models, while green dots show PPI+CC models.}
\label{fig:ertl}
\end{figure}

The iron core mass is typically below $\sim$\,$2.5\,M_\odot$ below the
PISN BH mass gap. The values for models above the PISN BH mass gap are
sensitive to the amount of nuclear burning going on before the
stopping criterion based on the infall velocity is reached, and
because of their large mass, the entropy is larger than $4k_BN_A$
throughout these stars. Because of the as of yet insufficient
understanding of black hole formation it is hard to predict whether
these models would give a successful, albeit possibly weak,
explosion. We would expect that if successful explosion can happen,
they would be powered by fallback accretion.

Figure~\ref{fig:ertl} shows our CC (blue) and PPI+CC (green) on the
plane used by \cite{ertl:16} to determine the ``explodability''. They
determine it using 1D parametric simulations aiming at reproducing
SN1987A starting with single star progenitors of initial mass
$15-20\,M_\odot$. The red area indicates the region where some of
their engines produce explosions and neutron star formation, while
others produce failed explosions and BHs (see their Tab.~2 and
Fig.~8). Above this (gray area), \cite{ertl:16} predict BH formation
with a failed explosion, while below they predict successful
explosions with neutron star formation. Applying this criterion to our
models requires extrapolating significantly outside the range
originally explored by \cite{ertl:16}, shown by the dotted rectangle
in the bottom left corner. We omit from the plot the models models
above the PISN BH mass gap in our grid for which $M_4$ and $\mu_4$ are
zero.  Most of our PPI+CC models fall in the failed explosion region
according to the \cite{ertl:16} criterion. A few enter in the region
that depends on the assumed calibration for their model, and some
cross marginally into the successful explosion and neutron star
formation region. The more explodable models correspond to lower
initial He core masses, but there is a large scatter in the trend. All
of our CC models fall in the region for BH formation without an
explosion.

\begin{figure}[htbp]
  \centering
  \includegraphics[width=0.5\textwidth]{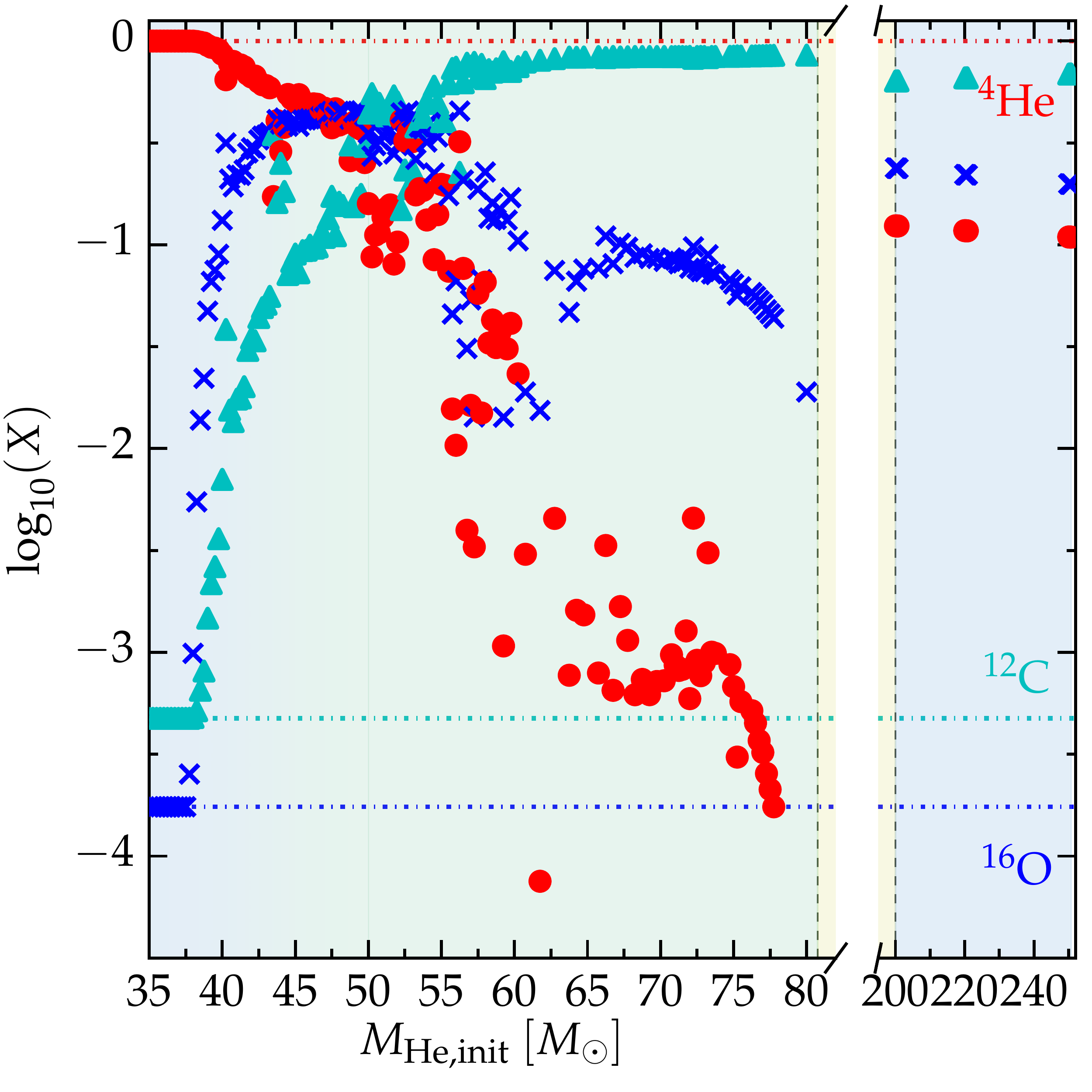}
  \caption{Surface composition at the onset of core-collapse for the PPI+CC and
    CC models. $^4\mathrm{He}$, $^{12}\mathrm{C}$, $^{16}\mathrm{O}$ are shown
    as red filled circles, cyan triangles, and  blue crosses, respectively. The dot-dashed
    horizontal lines of the same color mark the initial mass fraction for these
    elements for our $Z=0.001$ model grid. The colored background indicates the approximate evolution of the
    star (cf.~\Figref{fig:mhe_mbh} and \Secref{sec:overview}).}
  \label{fig:surf_comp}
\end{figure}

\Figref{fig:surf_comp} shows the surface mass fractions of $^4\mathrm{He}$,
$^{12}\mathrm{C}$,  $^{16}\mathrm{O}$ 
at the onset of core collapse for our CC and PPI+CC models. Should
the CSM evolve to be (partially) optically thin at the time of the
final CC and in case this results in a successful explosion, the
surface composition of the star and the mixing processes
\citep[e.g.,][]{dessart:12} would determine the spectral type
of the SN. Most PPI+CC models experiencing a significant amount of mass loss
show He-poor surfaces, and enhanced carbon (and to a smaller extent) oxygen mass
fractions, corresponding to type Ic SNe. Because of the radial expansion caused by the pulses, some of these
progenitors might look like extended and cool objects at the onset of collapse,
rather than compact and hot progenitors.
Conversely, if the He-rich CSM is optically thick, it might obscure the progenitor
star and the embedded explosion, and we expect that re-processing of the
photons by the shell would produce He lines (possibly narrow and in emission) corresponding
to a type Ib(n) SN.

We omit from \Figref{fig:surf_comp} the PISN models because our
stopping condition for these conservatively ensures the full star is unbound,
but it might not correspond to the ``beginning'' of the explosion. Nevertheless,
most PISN models start exploding while still retaining
some He at their surface (corresponding again to type Ib SNe) with the wind
mass-loss rates and metallicity adopted here.

\Figref{fig:surf_comp} also shows a trend with the initial He core mass: the
larger the initial $M_\mathrm{He,init}$, the more mass is lost to winds and PPI,
the lower the surface He mass fraction and the higher the carbon and oxygen mass
fractions. In the mass range
$50\,M_\odot\lesssim M_\mathrm{He,init}\lesssim 75\,M_\odot$, corresponding roughly to the region
where we find three distinct PPI-driven mass loss events, the predicted surface
abundances appear more noisy.

\section{Comparison to selected supernovae}
\label{sec:obs_SNe}

Stars experiencing the PPI+CC evolution should intrinsically be rare because of
the large initial mass necessary to build up a sufficiently massive core. To
produce a significant amount of CSM via PPI-driven pulses, our results suggest
that the He core mass needs to initially exceed
$M_\mathrm{He,init}\gtrsim42\,M_\odot$, corresponding to
$M_\mathrm{CO}\gtrsim 31\,M_\odot$.
Therefore, the rate of observed transients that can be interpreted as signatures
of PPI evolution should be small. Possibly for this reason an unambiguous detection of PPI+CC/PISN in
time-domain surveys is not yet available, although the physical mechanism underlying this
phenomenon is well understood. We consider here a few notable and
recent H-less type I SNe that have been proposed as PPI+CC candidates.

\paragraph{PTF12dam:}\cite{tolstov:17} modeled the H-less (type I) superluminous supernova PTF12dam
as powered by the combination of $^{56}\mathrm{Ni}$ decay and
CSM interaction. They proposed that
combination of energy sources invoking the following scenario: first the H-rich envelope
is removed by stellar winds, then the PPI pulses produce $\sim$\,$20-40\,M_\odot$
of CSM before the final CC synthesizes and ejects $M_\mathrm{^{56}Ni}\simeq6\,M_\odot$ of radioactive
material. Our results, albeit computed with a small nuclear reaction network,
never produce this combination of CSM mass and $M_\mathrm{^{56}Ni}$: PPI ejecta
exceeding $20\,M_\odot$ are found only for
$M_\mathrm{He,init}\gtrsim75\,M_\odot$ or equivalently
$M_\mathrm{CO}\gtrsim 51\,M_\odot$
(cf.~\Figref{fig:dm_pulses}), and only about $\sim0.2\,M_\odot$ of $^{56}\mathrm{Ni}$
is synthesized for PPI+CC models. Assuming that the final CC proceeds similarly
as for lower mass stars, we expect it would add $\sim$\,0.03-0.05$\,M_\odot$ of
$^{56}\mathrm{Ni}$ \citep[e.g.,][]{wongwathanarat:13}, which does not help to
reach the $M_\mathrm{^{56}Ni}$ claimed. An initial He core mass
exceeding  $M_\mathrm{He,init}\gtrsim140\,M_\odot$, or $M_\mathrm{CO}\gtrsim87\,M_\odot$, is required to
reach the amount of radioactive material required by \cite{tolstov:17}, which
would put the model in the PISN range where we do not expect CSM from PPI.

\paragraph{iPTF16eh:}\cite{lunnan:18} detected a time and frequency varying
Mg{\rm II} line in the spectrum of the type I superluminous supernova iPTF16eh.
They interpreted it as a light-echo of the explosion bouncing off a layer of
CSM at $r\simeq3.5\cdot10^{17}\,\mathrm{cm}$ moving at $\sim$\,$3\,300\,\kms$, implying an ejection
$\sim$\,$30\,\mathrm{years}$ before the final CC. Based on these CSM
properties and the models from \cite{woosley:17}, they inferred a progenitor
with $M_\mathrm{He,init}\simeq50-55\,M_\odot$ (or equivalently an initial total mass
$\sim$\,115$\,M_\odot$). Our models are in overall agreement with the results from
\cite{woosley:17} used by \cite{lunnan:18} to interpret iPTF16eh, although the
delay time and ejecta velocity would agree better with a slightly more massive
progenitor, with $M_\mathrm{He,init}\simeq60-65\,M_\odot$, i.e., $M_\mathrm{CO}\simeq43-46\,M_\odot$.  

\paragraph{SN2016iet:}\cite{gomez:19} analyzed the double-peaked peculiar type
I SN~2016iet. They explored several scenarios (PISN, CSM interaction,
and central engine) to power its light curve. This event showed an
unusually high Ca/O ratio, and extreme offset from the nearest galaxy
of $\sim$\,16\,kpc, however H$\alpha$ lines appear in the spectra
beyond 400 days, possibly indicating local star formation
activity. They also detected a possible light-echo from a H- and
He-poor shell moving at few thousand $\kms$. Regardless of the
scenario assumed, they inferred a large progenitor mass with a CO core
$55\,M_\odot\lesssim M_\mathrm{CO}\lesssim 120\,M_\odot$. The model
they favor to explain the light curve combines the signal from the
shock cooling of the prompt explosion (first peak) and CSM
interactions (second peak), but requires $\sim$\,$35\,M_\odot$ of CSM,
although this value is considerably uncertain (Gomez et
  al.~private communication).  Both the inferred presence of a shell
of H- and He-poor material and the claimed progenitor and CSM masses
suggest PPI+CC as a viable scenario for the progenitor of
SN2016iet. Several models with initial He core mass
$M_\mathrm{He,init}\gtrsim50\,M_\odot$ produce PPI-driven pulses with
mass, timing, and velocity within a factor of about two from the
values inferred by \cite{gomez:19}.  However, reaching that total
amount of CSM would either require the progenitor to be at the very
edge of the PISN regime, for which we find long interpulse delays
(cf.~\Figref{fig:timing}) and also the last pulse tends to produce
little mass loss (cf.~\Figref{fig:dm_pulses}). Alternatively,
relaxing the requirement to match the total CSM mass budget by
allowing for a contribution of the stellar wind to
the CSM (e.g., because of the wind in between pulses running into a
slower-moving previously-ejected shell), models with
$60\,M_\odot \lesssim M_\mathrm{He,init}\lesssim 70\,M_\odot$, i.e.,
$43\,M_\odot \lesssim M_\mathrm{CO} \lesssim 49\,M_\odot$, produce
pulses removing larger amounts of mass in the final few years of the
progenitor's life. This might produce a better agreement with the
observed features of SN2016iet. If that were the case, this event
might be the birth of one of the most massive BHs predicted below the
PISN mass gap, cf.~\Figref{fig:mhe_mbh}. At
\href{https://doi.org/10.5281/zenodo.3406356}{https://doi.org/10.5281/zenodo.3406356},
we provide models computed at our fiducial metallicity value and at
the metallicity of the galaxy at $16\,$kpc from SN2016iet
($Z=0.00198\simeq0.14Z_\odot$), although it is likely that a dimmer
galaxy, possibly with different $Z$, is coincident with and presently
outshined by SN itself. These can provide input for more detailed
calculations of the CSM structure needed to compare with SN2016iet.

\paragraph{SN2006jc, PS15dpn and other narrow-line SNe:}~Two out of
the three SNe we considered above are super-luminous, however the
final collapse of a PPI+CC progenitor or PISNe does not need to be
superluminous \citep[][]{woosley:17}. The PPI is just one possible
mechanism to create CSM, which can produce extreme luminosities by
generating radiation from the kinetic energy of the ejecta and/or
narrow emission lines (even if the luminosity does not reach extreme
values). The detection of narrow H lines determines the classification
of a SN as a type IIn, while the detection of narrow He emission lines
determines the classification as type Ibn. Both kinds of event are too
common to be entirely explained with PPI+CC progenitors, and it is
likely that both classes contain events with a diversity of physical
mechanisms (e.g.,~\citealt{pastorello:08} but see also
\citealt{hosseinzadeh:17}). Nevertheless, it is possible that at least
some of these events might correspond to the observational counterpart
of the death of PPI+CC progenitors. In particular, our simulations can
produce several solar masses of H-free CSM moving at a few thousand
$\kms$, which correspond to the width of the He lines detected in some
SN~Ibn without any fine-tuning required. Even if the detection of narrow
  lines alone is not sufficient to associate a specific SN to a PPI event,
  combining evidences from previous coincident transients, large
  ejecta masses or long lightcurve durations, 
  large $^{56}\mathrm{Ni}$ yields, an extremely young surrounding stellar population,
  and/or nucleosynthetic signatures might strengthen the case for
  associating specific event with this scenario.  Possible examples of SN~Ibn
that might correspond to PPI+CC are SN2006jc and PS15dpn. The former
showed relatively narrow He lines possibly hinting to asphericity of
the CSM \citep{foley:07} and was spatially coincident with an
unexplained outburst two years earlier
\citep[e.g.,][]{pastorello:07,foley:07}. For the latter,
\cite{wang:19} proposed to fit the light curve by combining CSM interaction and radioactive decay,
and inferred CSM and $^{56}\mathrm{Ni}$ masses of
$\sim$$0.8\,M_\odot$ and $\sim$\,$0.1\,M_\odot$, respectively, in good agreement with our models.

\section{Limitations and caveats}
\label{sec:caveats}

The stellar evolution simulations presented here require a large number of
assumptions. Work to assess the robustness of these calculations has
been carried out by \cite{marchant:19,
  farmer:19, renzo:20:conv_PPI} (see also Appendix~\ref{sec:res_study}), to which we refer the readers for more details.

\subsection{Ignition location and spherical symmetry}
\label{sec:ignition_location}

\begin{figure}[tbp]
  \centering
  \includegraphics[width=0.5\textwidth]{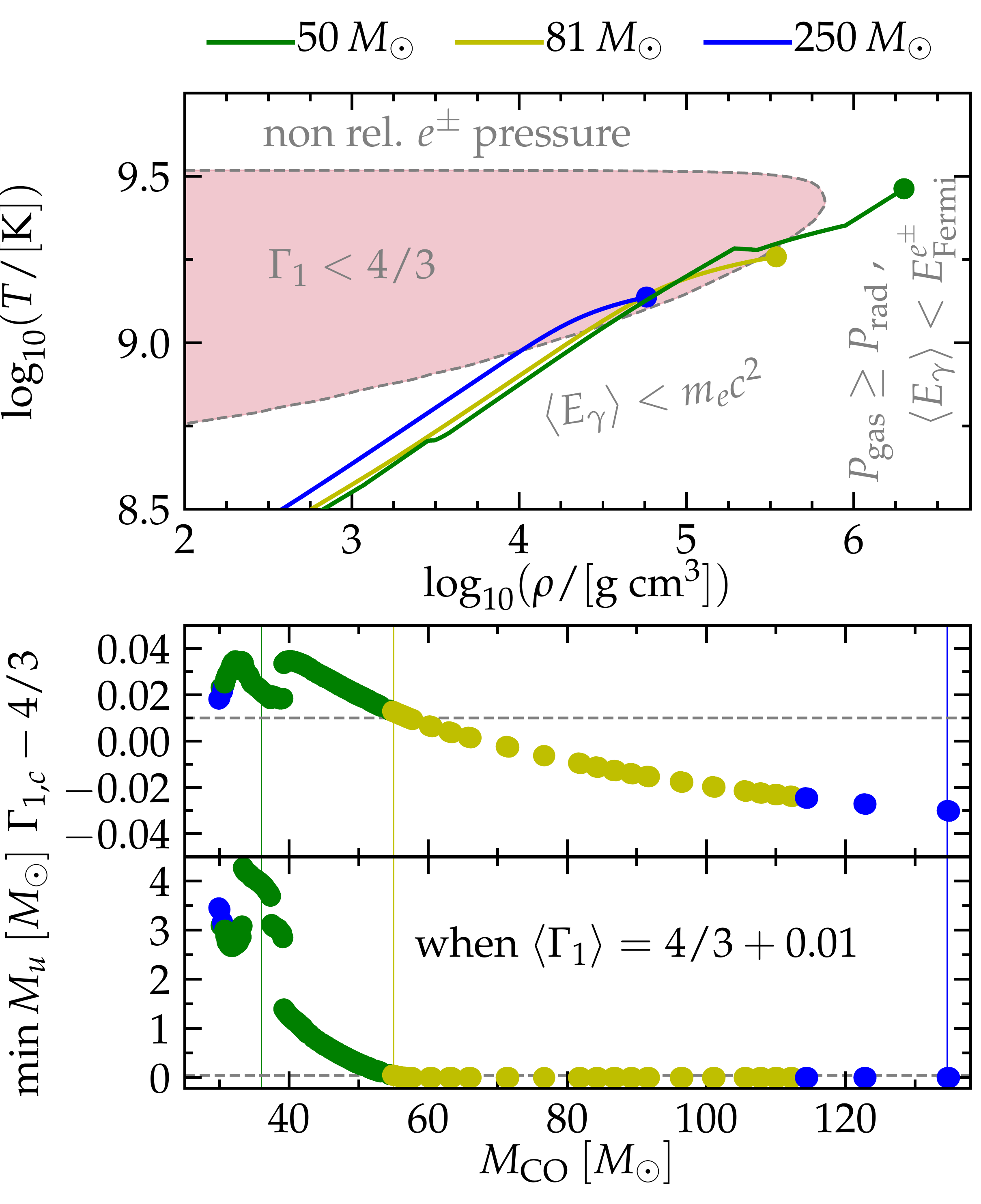}
  \caption{Top panel.~Temperature and density profiles for example models approaching the
    instability, i.e., the first time
    $\langle\Gamma_1\rangle-4/3=0.01$. The filled circles mark the
    central conditions at this stage. The green, yellow, and blue lines show
    examples of PPI+CC, PISN, and CC above the mass gap, respectively. All three examples are labeled according
    to their $M_\mathrm{He,init}$. The red area
    marks the region of the EOS where pair-production results in an (local) instability. Bottom panel.~Values of the adiabatic index in the
    center when approaching the instability for the entire
    grid. The thin vertical lines mark the $M_\mathrm{CO}$ of the
    examples models of the corresponding color in the top panel.}
  \label{fig:EOS_grid}
\end{figure}

One of the key assumptions is that spherical symmetry is maintained during the
evolution. \cite{chen:14,chen:19} showed that if a
pulse starts symmetrically, hydrodynamic instabilities only weakly deform the
pulse. However, the first stellar layer to become unstable due to pair
production in a pulsating model is not
necessarily at the very center, especially at the lower
mass end of the PPI+CC regime. The top panel of \Figref{fig:EOS_grid} shows the
temperature and density profile of three examples with
$M_\mathrm{He,init}=50,\,81,\,250\,M_\odot$, representative of PPI+CC, PISN, and CC
above the mass gap, respectively. These $M_\mathrm{He, init}$
correspond to $M_\mathrm{CO}=36.04,\,55.02,\,134.60\,M_\odot$, respectively. The stellar tracks are plotted at the time when the
volumetric pressure-weighted average $\langle\Gamma_1\rangle$ first approaches the instability
value 4/3, i.e., $\langle\Gamma_1\rangle-4/3=0.01$. The red shade emphasizes the instability region
(neglecting its weak dependence on the details of the chemical composition), and
the text annotations indicate the physical ingredients that stabilize the
structure outside of this region \citep[][]{zeldovich:99, kippenhahn:13}.

The middle panel of \Figref{fig:EOS_grid} shows the local value of the
adiabatic index in the center $\Gamma_{1,c}$ across our grid, also
plotted when each model first reaches
$\langle\Gamma_1\rangle-4/3=0.01$. The bottom panel of
  \Figref{fig:EOS_grid} shows the innermost unstable mass coordinate
  $M_u$, i.e.~the innermost location where $\Gamma_1 <4/3$. The
colors in the middle and bottom panel have the same meaning as
in \Figref{fig:mhe_mbh}. Overall, as $M_\mathrm{CO}$ increases,
the central value of the adiabatic index $\Gamma_{1,c}$ at the
beginning of the instability decreases, and the location of the
instability moves inward. However, this trend is not completely
monotonic in the central part of the PPI+CC regime.

Two example models that ultimately result in a core collapse are shown in the top
panel of \Figref{fig:EOS_grid}. The green line corresponds to our
$M_\mathrm{He,init}=50\,M_\odot$ example for PPI+CC, which shows more features
compared to the other models, because the chemical stratification is more
important in lower mass models. The central region of the $50\,M_\odot$ model
has a local value of the adiabatic index $\Gamma_{1,c}-4/3>0.01$ when $\langle\Gamma_1\rangle\simeq4/3$, i.e., the center is stable when the
averaged $\langle\Gamma_1\rangle$ approaches instability. The deepest interior
is too dense to become unstable: the contribution of radiation
pressure to the total pressure decreases and the $e^{\pm}$ pairs fill the available continuum energy levels,
raising the Fermi energy $E^{e^\pm}_\mathrm{Fermi}$ and consequently the minimum
energy photons need to produce a pair, preventing layers from undergoing the
runaway instability \citep[e.g.,][]{zeldovich:99}. This is true for all our PPI+CC models: in
the bottom panel of \Figref{fig:EOS_grid}, all the green points
corresponding to PPI+CC have central values $\Gamma_{1,c}-4/3>0.01$
when $\langle\Gamma_1\rangle-4/3=0.01$.

Since all our models are strongly radiation pressure dominated, they
evolve with most of the mass along the $\Gamma_1\simeq 4/3$ locus on
the $(\rho,T)$-plane, and when the instability starts, a nearly
homologous contraction ensues. However, in the densest part of the
core, neutrino cooling dominates over the energy release of the
burning. Therefore, the net energy release starts off-center in our
$50\,M_\odot$ example, and because of the assumption of spherical
symmetry, the energy release occupies a spherical mass shell. However, in
nature the ignition might not happen simultaneously across the entire
spherical shell, and this could potentially seed an asymmetric
explosion. If asymmetries can build up rapidly during the
pair-instability driven explosion (possibly aided by rotation), they
could also lead to orbital ``kicks'' when PPI happens in a binary
\citep[][]{marchant:19}.

The location of the unstable layer  at the onset of the instability and of the location where the
instability triggers net energy release both move inward toward
the center as $M_\mathrm{He,init}$ and $M_\mathrm{CO}$ increase. The least massive models to go
PISN are characterized by having their center close to the instability,  i.e.,
$\Gamma_{1,c}-4/3\leq0.01$, when $\langle\Gamma_1\rangle-4/3=0.01$ for
the first time, as shown by the yellow filled circles
in the bottom panel and the yellow solid line corresponding to
$M_\mathrm{He,init}=81\,M_\odot$ in the top panel of \Figref{fig:EOS_grid}. Our
least massive PISN model has $M_\mathrm{He,init}=80.75\,M_\odot$
(corresponding to $M_\mathrm{CO}=54.89\,M_\odot$) and a central value 
$\Gamma_{1,c}-4/3=0.01$ at this evolutionary stage. This model
has $M_u\simeq0.05\,M_\odot$ (shown as a dashed horizontal line in the bottom
panel of \Figref{fig:EOS_grid}), i.e.~the unstable layer extends almost all the
  way to the center.

Models forming a BH above the PISN mass gap (cf.~blue line in the top panel of
\Figref{fig:EOS_grid} for a $M_\mathrm{He,init}=250\,M_\odot$ He core) are also unstable in their very center when $\langle\Gamma_1\rangle$
reaches 4/3, but the ensuing thermonuclear explosion does not cause
either pulses or full disruption. The different outcome is not caused by lack of energy released in the explosions, but rather
by the inefficient use of that energy \citep[][]{bond:84}.

To summarize, the temperature, composition, and density profile of the star
when it approaches the instability (i.e., $\langle\Gamma_1\rangle-4/3=0.01$ for the first time) are indicative of its future
evolution. In particular, the
local value of the adiabatic index in the center $\Gamma_{1,c}$ at this point can be
used to approximately distinguish PISN evolution with no BH remnant (if also
$\Gamma_{1,c}-4/3\leq0.01$) or PPI+CC evolution with a BH remnant (if instead the center
is safely stable with $\Gamma_{1,c}-4/3>0.01$ when the star as a whole is becoming unstable). This provides a
criterion to decide the final fate of a star without having to compute the
hydrodynamical phase.

\subsection{CSM structure and composition}

We described in \Secref{sec:grid_ejecta} the amount of mass ejected, its initial
velocity, and the ejection timing resulting from our simulations. We typically keep the ejecta on our Lagrangian grid
for several timesteps after ejection (until either the bound layers have recovered hydrostatic
equilibrium or the onset of CC is reached). These ejected layers are moving significantly faster than the
escape velocity and the sound speed, and our PPI+CC models exhibit an overall
velocity gradient increasing outward¸ so the ejecta do not cause any back-reaction on
the inner layers that remain bound. Based on the initial
mass, velocity, and time of the ejections, \Secref{sec:grid_ejecta} illustrates the main features
we expect in the CSM structure surrounding these stars with a toy model
assuming propagation at constant velocity of
the ejecta. This is an oversimplification, since the low density ejecta are
likely to be optically thin and thus can lose energy radiatively. Moreover, as
previously noted by \cite{woosley:17}, the ejected shells can in many cases
collide with each other, and this could also significantly change the CSM
structure at the end. Multidimensional radiation hydrodynamics calculations using our
results as input for the mass, chemical composition, and thermal state
of the ejecta could be used to predict more robustly the CSM structure around
PPI+CC models for comparison with observed transients, and to address the
question of how many progenitors might reach the final CC embedded in
a optically-thick layer of previously ejected material.

Another assumption in our calculations is that the presence of a H-rich
envelope can be neglected to study the dynamics of PPI \citep[][]{woosley:17, woosley:19}
and that, even if present, such a envelope would be removed early in the
evolution by winds or binary interactions. Should a star retain some H-rich material until the onset of the
first pulse, we can estimate if it would be detectable in the
CSM surrounding these stars assuming that (\emph{i}) the PPI-driven mass loss timing is
unaffected by the presence of the H-rich envelope and (ii) the entire
envelope is ejected in the first pulse \citep[][]{woosley:17}.

To estimate the ejection velocities and radii of the H-rich material,
we ran a H-rich $140\,M_\odot$ model with initial He abundance
$Y=0.27$ and metallicity $Z=0.001$ with the same setup as in our
grids. In Appendix~\ref{sec:hrichcomp}, we compare this model to a He
core of $55.25\,M_\odot$ which produces a
$M_\mathrm{CO}\simeq39.41\,M_\odot$, similar in mass to the CO core
produced by the H-rich $140\,M_\odot$ star. The H-rich model reaches the
onset of the PPI ($\langle \Gamma_1\rangle-4/3=0.01$) with
$\Gamma_{1,c}-4/3=0.05$, so we expect it to follow the PPI+CC evolutionary path based
on \Secref{sec:ignition_location}. This expectation is
confirmed by our results presented in Appendix~\ref{sec:hrichcomp}. At
the onset of the instability and with our assumed wind mass loss, this model has
a total mass of $M_\mathrm{tot}=83.2\,M_\odot$, a He core of
$M_\mathrm{He}\simeq50\,M_\odot$, and a CO core of
$M_\mathrm{CO}\simeq39\,M_\odot$. The remaining envelope has a mass of
$ M_\mathrm{env}\equiv M_\mathrm{tot}-M_\mathrm{He}\simeq33\,M_\odot$, however, the composition
of this envelope is dominated by He, with a H mass fraction of
$X\simeq0.14$, since the winds have carved out material down to the initial location
of the main sequence core of the star. At this stage, the envelope spans from
the He core edge at $R_\mathrm{He\,core}=0.58\,R_\odot$ to $R_*=3637\,R_\odot$, so it is
significantly extended.

If we assume propagation at constant velocity of this envelope, we can estimate the minimum and maximum
radii ($R_\mathrm{min}$ and $R_\mathrm{max}$, respectively) of this
H-rich material at the time of the final CC and its average density
($\langle \rho\rangle$) with

\begin{align}
\label{eq:estimates}
\begin{split}
  R_\mathrm{min} = R_\mathrm{He\,core}+v_\mathrm{esc,in}\times(t_\mathrm{CC}-t_\mathrm{pulse\ end}) \ \,\\
  R_\mathrm{max} = R_*+v_\mathrm{esc,out}\times(t_\mathrm{CC}-t_\mathrm{pulse\ end}) \ \,\\
  \langle\rho\rangle = \frac{3 M_\mathrm{env}}{4\pi (|R_\mathrm{max}^3-R_\mathrm{min}^3|)} \ \ ,
\end{split}
\end{align}

where $t_\mathrm{CC}-t_\mathrm{pulse\ end}\simeq10\,\mathrm{years}$ based on the
$M_\mathrm{He}$ and $M_\mathrm{CO}$ of this model and \Figref{fig:timing}. If we
assume both escape velocities to be the stellar surface escape
velocity, i.e., $v_\mathrm{esc,out}=v_\mathrm{esc,in}=\sqrt{2G M_\mathrm{tot}/R_*}$, then
the CSM layer containing H would be at
$R_\mathrm{min}\simeq10^{15.46}\,\mathrm{cm}\lesssim r \lesssim R_\mathrm{max} \simeq\,10^{15.51}\,\mathrm{cm}$ and have an average density of
$\langle \rho\rangle\simeq2.2\cdot10^{-12}\,\mathrm{g\ cm^{-3}}$.

Instead, if we assume,
$v_\mathrm{esc,in}=\sqrt{2G M_\mathrm{He}/R_\mathrm{He\,core}}$ is the velocity
from the He core edge,  while keeping
the same $v_\mathrm{esc,out}$, then we obtain
$R_\mathrm{max}<R_\mathrm{min}$ (meaning there would necessarily be collisions
internal to the ejecta that our simplistic toy model ignores). Nevertheless, the
minimum radius at which we would expect CSM with H in it form the PPI
mass ejection is $R_\mathrm{max}\simeq3.2\cdot10^{16}\,\mathrm{cm}$ and such layer would
have an average density of
$\langle\rho\rangle\simeq3\cdot10^{-18}\,\mathrm{g\ cm^{-3}}$. The
large difference between these two estimates is mostly due to the
large radial extent of the envelope.

\begin{figure*}[htbp!]
  \centering
  \includegraphics[width=\textwidth]{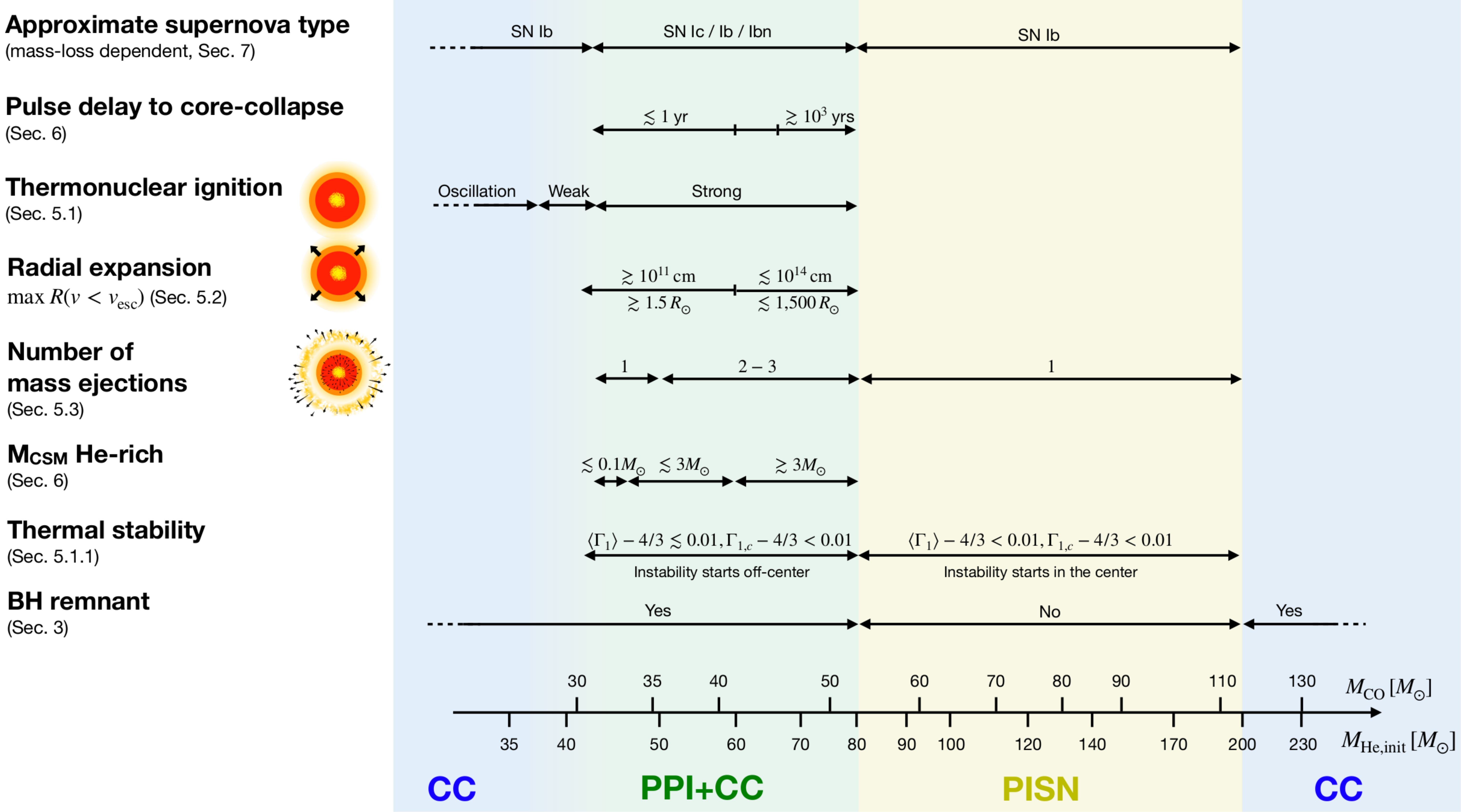}
  \caption{Summary of the pair-instability driven behavior of models as a
    function of their initial He core mass ($M_\mathrm{He,init}$) and
    maximum carbon-oxygen core mass ($M_\mathrm{CO}$). The
    approximate SN type in the top row speculates on what the 
    resulting SN would appear like, if the
      final collapse produces an explosion.}
  \label{fig:summary}
\end{figure*}

\section{Conclusions}
\label{sec:conclusions}

The broad theoretical understanding of the predicted pair-instability driven
transients has been well established for several decades, however they remain somewhat
elusive from an observational perspective. Recent developments in stellar
evolution calculations allow for the exploration of synergies between
gravitational waves and time-domain observations to better understand the
formation process of the most massive stellar BHs. 

We have computed a grid of naked He star models in the mass range
$35\,M_\odot\leq M_\mathrm{He,init} \lesssim 250\,M_\odot$ to investigate whether these would
experience phases of global dynamical instability and pulsational mass loss due
to the pair-production instability. We have computed grids at two different
metallicities, $Z=0.001$ and $Z=0.00198=14\%\,Z_\odot$, although the main features we discuss are
not significantly dependent on $Z$ (except for the wind mass loss rate, see also
\citealt{farmer:19}). All our
input files and numerical results are available at
\href{https://doi.org/10.5281/zenodo.3406356}{https://doi.org/10.5281/zenodo.3406356}.

\Figref{fig:summary} summarizes our main results across the mass range
considered. We find, in agreement with previous studies, that stars enter into the
PPI regime progressively. The production of $e^{\pm}$ initially causes
``oscillations'' of the core temperature and nuclear luminosity at the lowest
mass end. The least massive models experiencing an explosive thermonuclear ignition
($M_\mathrm{He,init}\gtrsim37.5\,M_\odot$) do not suffer significant global consequences
(``weak pulses''). Increasing further the initial He core mass, the pulses become
progressively stronger, causing at first large radial expansions (for initial
$41\lesssim M_\mathrm{He,init}\lesssim42\,M_\odot$), and finally (for initial
$M_\mathrm{He,init}\gtrsim42\,M_\odot$) also the ejection of matter. The values quoted
here are for the initial He core mass of our models, which can be interpreted as
the core mass at the end of the main sequence of the star. The mapping of these values
to the final (preinstability) He core mass is mass loss and metallicity
dependent (see also \citealt{farmer:19}).

The different effects of a pulsational pair instability event on the star allow
for (at least) three different physically-motivated definitions of a ``pulse'', depending on
which observable is considered (\Secref{sec:pulses_def}). The number
of pulses and which He and CO core masses produce pulses might
vary significantly depending on the observable of interest.

The first definition of pulse we consider (\Secref{sec:thermo_def}) is based on the core
thermonuclear ignition, following the historical development of studies of
pair-instability evolution. We find that in the lowest mass models the core
ignition does not produce an observable electromagnetic signal or a significant
impact on the final BH mass: the nuclear energy released in the burning is
redistributed and stored in the star without affecting significantly the
outermost layers. The most promising way to detect directly these
core-ignition events is through the variations in the neutrino luminosity.

The second definition is based on the radial expansion of the models in response
to the core ignition (\Secref{sec:radial_def}): this definition shifts
the lower edge of the pulsating regime upward
in mass. For the most massive
pulsating models (which also eject mass), the radial expansion itself might be
hidden behind a pseudo-photosphere in the ejected layers. Because of the rarity
of these stars in the local Universe, the most promising way
to detect these radius variations may be through their enhancement of the rate of binary
interactions \citep[e.g.,][]{marchant:19}, although nearby stars that
might undergo this evolution in the future exist
\citep[e.g.,][]{bestenlehner:11, crowther:16, renzo:19:vfts682}.

The third definition is based on the ejection of mass
(\Secref{sec:mejection_def}), which impacts both the circumstellar material
around these stars and the final BH mass they produce. The ejected matter
creates shells of ejecta surrounding the star. If the final collapse results in
a successful (even if weak) explosion, the final SN ejecta can hit this
PPI-produced CSM and convert kinetic energy into radiation.

In our grid, we find full disruption in a PISN for an initial He core mass of
$M_\mathrm{He,init}\simeq80\,M_\odot$ corresponding to a final He core mass of
$M_\mathrm{He}\simeq60\,M_\odot$ and $M_\mathrm{CO}=55\,M_\odot$ after the wind mass loss.
With our assumptions for the wind mass loss and metallicity, most PISN models would still retain
He-rich material at their surface at the onset of the explosion. We propose a
simplified criterion to distinguish full disruption in a PISN from pulsational
behavior producing a final BH based on the adiabatic index at the center of the
star $\Gamma_{1,c}\lesssim0.01$ at the onset of the instability, defined as the first moment when
the volumetric pressure-weighted average of the adiabatic index
$\langle\Gamma_1\rangle-4/3=0.01$
(\Secref{sec:ignition_location}).
This threshold value
allows for the approximately estimate the fate of a stellar model
without the need to compute the hydrodynamical evolution.

The signature on the final BH masses is potentially detectable with a population of gravitational wave sources
\citep[][]{fishbach:17, talbot:18, stevenson:19, mangiagli:19}. Only for
$M_\mathrm{He,init}\gtrsim42\,M_\odot$
($M_\mathrm{CO}\simeq30.75 \,M_\odot$) is the stellar core mass significantly reduced and
the circumstellar material significantly affected, as shown in \Figref{fig:summary}.
The maximum BH mass below the PISN mass gap that we find is $\sim$\,$45\,M_\odot$
and it is formed by the collapse of an initially
$M_\mathrm{He,init}\simeq62\,M_\odot$ ($M_\mathrm{CO}\simeq43.5\,M_\odot$) He core that
went through pulsational mass loss. More massive He cores also produce BHs, but
because of the stronger mass loss due to winds and pair-instability driven
pulses, the resulting BH masses are smaller.

Pair instability does not result in full disruption of the star for an
initial He core mass of $M_\mathrm{He,init}\simeq200\,M_\odot$, which forms a BH of mass
$M_\mathrm{BH}=125\,M_\odot$ after wind mass loss. Above this He core mass, the
photodisintegration of newly synthesized heavy elements during the thermonuclear
explosion prevents the disruption of the entire star \citep[e.g.,][]{bond:84}. The boundaries between
PISN and BH formation we find are in very good agreement with
previously published results.

We have characterized the CSM properties around the pulsating models resulting
in mass ejections by assuming unperturbed constant velocity propagation of the
ejecta. Under this simplifying assumption, we find that the CSM mass grows
almost monotonically with the initial $M_\mathrm{He,init}$ from $\sim$\,$10^{-6}\,M_\odot$ (for
$M_\mathrm{He,init}\simeq42\,M_\odot$) to $\sim$\,$20\,M_\odot$ at the edge of the PISN range.  For
initial $M_\mathrm{He,init}\gtrsim50\,M_\odot$, the combined ejection of matter and mixing
during a pulse propagation make He less abundant than C and O at the
stellar surface at the onset of the final core-collapse. The stellar surface
at the onset of core collapse might be obscured by the previously ejected layers.

The velocity of the ejecta is a few thousand $\kms$, with the first mass-loss
event often resulting in larger velocities. Nevertheless, with our assumptions,
we find numerous self-collisions of the ejecta with previously ejected layers in
agreement with the predictions of \cite{woosley:07} and \cite{woosley:17}. This velocity range is
close to the width of narrow He lines detected in some SN~Ibn.

The timing of pair-instability driven mass ejections also spans a large
range of values, with a systematic trend of longer delays between pulses for the more massive
models. This is because more massive models produce more energetic pulses that
require a longer time (up to $\sim$\,$10^4$ years) for the star to recover its equilibrium.

With the velocity and timing of the ejecta produced by our models, we expect
the PPI-produced circumstellar material to be at
$\sim$\,$10^{12}-10^{16}\,\mathrm{cm}$ away from the collapsing star at the end of
its evolution. This range covers the distances inferred in observational
candidates for pulsational pair instability evolution. 

Gravitational wave detections of merging binary black holes are
rapidly accumulating during the third LIGO/Virgo observing run, and
currently available constraints on their mass will soon become
statistically stringent. Together with the ongoing and upcoming large
time-domain survey which will reveal a plethora of transient,
including rare and exotic ones, this will provide direct constraints
on pair-instability evolution of the most massive stars. Thus, in the
future, gravitational and transient observations will soon shed
light on the pair-instability evolution of the most massive stars and
the BHs these produce.

\begin{acknowledgements} We are grateful for the in depth reading of
  the referee and the suggested improvements to this manuscript. We acknowledge helpful
  discussions with M.~Cantiello, D.~Hendricks, E.~Laplace, I.~Mandel, B.~Paxton,
  F.~Timmes, L.~van~Son, and A.~Vigna-G\'omez. MR, SJ, and SdM
  acknowledge funding by the European Union's Horizon 2020 research
  and innovation programme from the European Research Council (ERC)
  (Grant agreement No.\ 715063), and by the Netherlands Organisation
  for Scientific Research (NWO) as part of the Vidi research program
  BinWaves with project number 639.042.728.  RF is supported by the
  Netherlands Organisation for Scientific Research (NWO) through a top
  module 2 grant with project number 614.001.501 (PI de Mink). EZ
  acknowledges support from the the Swiss National Science Foundation
  Professorship grant (project number PP00P2 176868) and from the
  Federal Commission for Scholarships for Foreign Students for the
  Swiss Gov- ernment Excellence Scholarship (ESKAS No. 2019.0091) for
  the academic year 2019-2020. Simulations
  were carried out on the Dutch national e-infrastructure (Cartesius,
  project number 16343) with the support of the SURF Cooperative.
\end{acknowledgements}

\bibliographystyle{aa}
\bibliography{./bibliography/ppisn}
\appendix

\section{Resolution study}
\label{sec:res_study}

\begin{figure}[tbp!]
  \centering
  \includegraphics[width=0.5\textwidth]{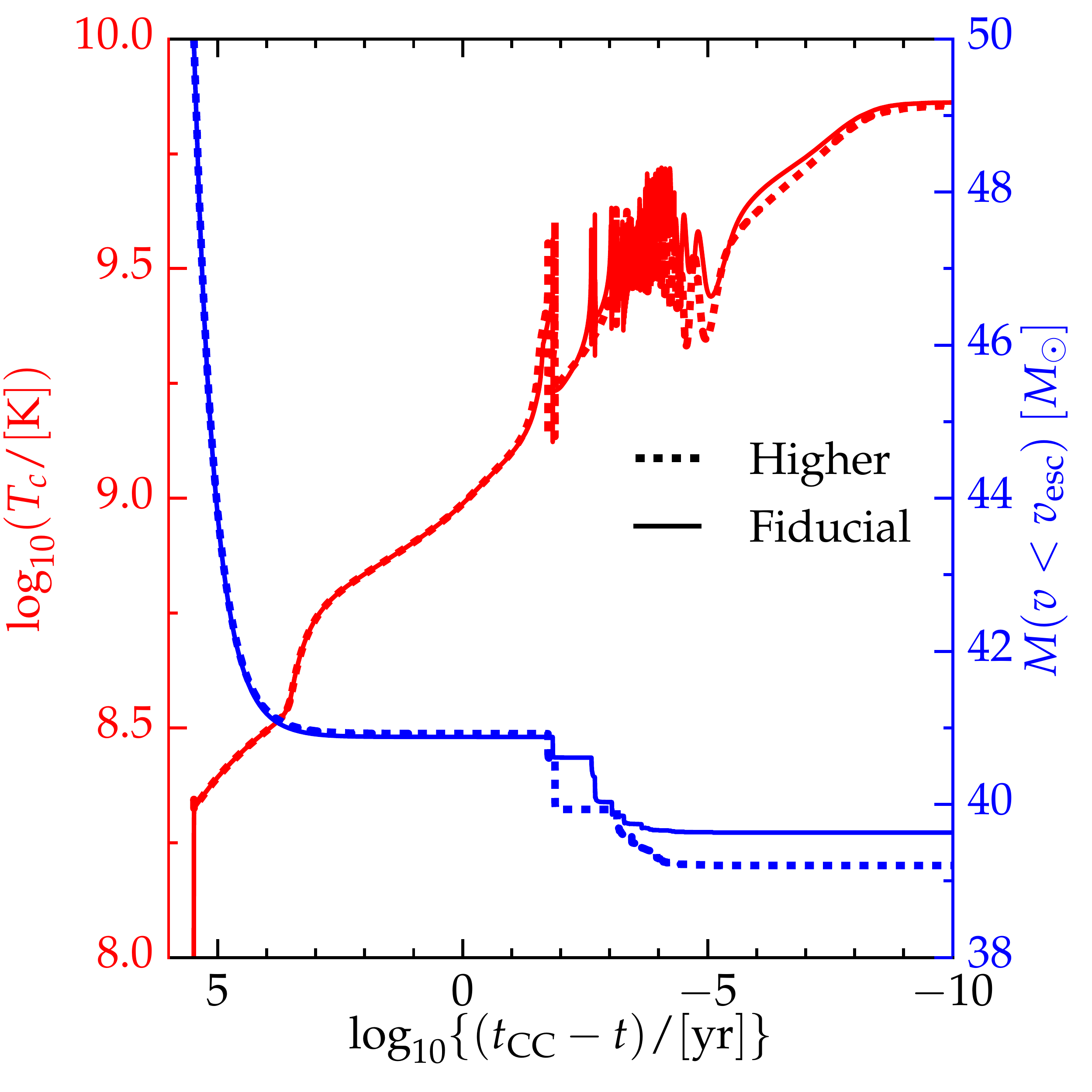}
  \caption{Mass (blue, right axis) and central temperature (red, left axis) evolution for our example $50\,M_\odot$
    example He core. The scale on the time axis emphasizes the short lived final
    phases. The differences in the final mass due to variations in the spatial
    and temporal mesh arise mostly from the dynamical phase of evolution and are
    limited to $\Delta M\lesssim0.5\,M_\odot$.}
  \label{fig:res_study}
\end{figure}

We present here a study of the impact of the numerical resolution in Lagrangian
mass coordinate and time on our results. We refer the interested readers to
\cite{marchant:19} for a study on the relaxation procedure and of the
numerical resolution on the pulse mass loss, to 
\cite{farmer:19} for a more comprehensive study of the impact of the numerical
resolution and input physics variations on our PPI models, and
\cite{renzo:20:conv_PPI} for a study of the impact of different treatments of
time-dependent convection.

\Figref{fig:res_study} shows the evolution in time of the amount of mass bound
to the star (blue) and its core temperature (red) for two $50.0\,M_\odot$ He core
models computed with different resolutions. \texttt{MESA} offers many controls
to fine-tune the resolution (see also the provided \texttt{inlist}s and
\texttt{run\_star\_extras.f}), here we vary only three parameters governing the
overall variations of averaged quantities across adjacent mesh points and across
timesteps. Our fiducial (higher) resolution uses
\texttt{mesh\_delta\_coeff}=\texttt{0.8} (\texttt{0.6}) and
\texttt{split\_merge\_amr\_nz\_baseline}=\texttt{8\,000} (\texttt{10\,000}) for
the spatial resolution during the hydrostatic and hydrodynamical phases of
evolution, respectively. The time discretization is controlled through
\texttt{varcontrol\_target}=\texttt{5d-5} (\texttt{1d-5}). The largest
differences in the evolution are found after the onset of the PPI pulses, during
the dynamical phase after
$\log_{10}\{(t_\mathrm{CC}-t)/\mathrm{[yr]}\}\lesssim-2$. These result in a
$\Delta M=0.43\,M_\odot$ difference in the mass remaining bound (and on the
amount of mass ejected). We emphasize that even our fiducial value provides a
resolution significantly higher than the MESA defaults, with a number of mesh
points $1289\lesssim N\lesssim 6311$ and 87900 timesteps from the onset of He core burning to
the onset of core-collapse. For comparison, the higher-resolution model shown in
\Figref{fig:res_study} has $1583\lesssim N\lesssim 7908$, however it is able to finish the
evolution using a slightly smaller number of timesteps, 86165. This likely
indicates that at the higher spatial resolution the most stringent condition on
the timesteps is not \texttt{varcontrol\_target}, and that the higher spatial
resolution provides more numerical stability of the solution allowing for
longer timesteps.

\section{Comparison to H-rich model}
\label{sec:hrichcomp}

A full characterization of PPI+CC evolution for stars with a H-rich
envelope is beyond the scope of this study (we refer interested
readers to \citealt{woosley:17, leung:19}). Here, we present a brief
comparison between a $140\,M_\odot$ H-rich model (computed assuming
and initial He abundance of $Y=0.27$) and a
$M_\mathrm{He, init}=55.25\,M_\odot$. Both these models produce
$M_\mathrm{CO}\simeq40\,M_\odot$.

\begin{figure}[bp!]
  \centering
  \includegraphics[width=0.5\textwidth]{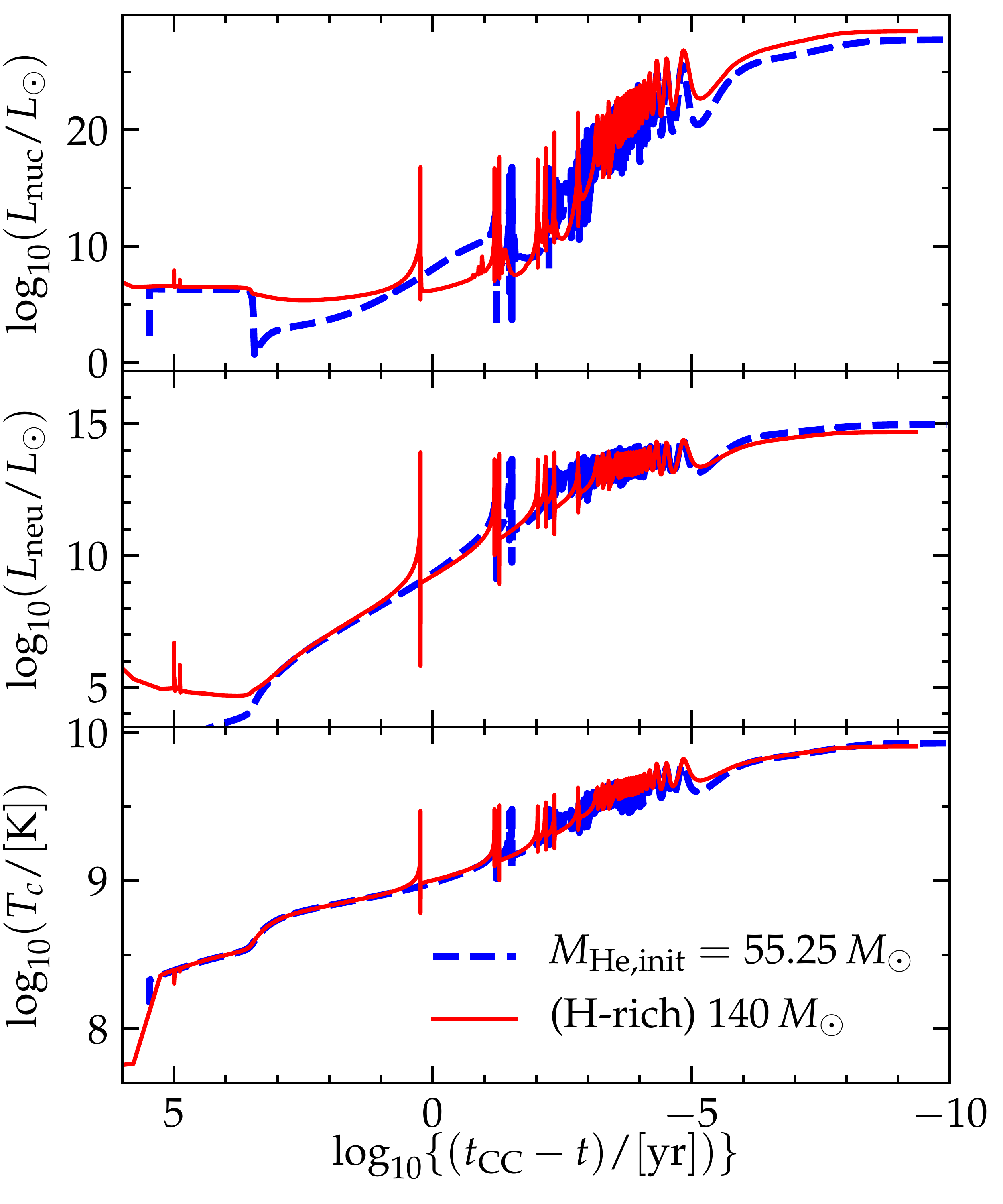}
  \caption{Comparison of the evolution of the nuclear
    luminosity (top) and neutrino luminosity (center) integrated
    throughout the star, and central temperature (bottom) for a H-rich initially $140\,M_\odot$
    star (red solid lines) to a $M_\mathrm{He,init}=55.25\,M_\odot$
    core (blue dashed lines). Both models form a $M_\mathrm{CO}\simeq40\,M_\odot$.}
  \label{fig:core_comparison}
\end{figure}

\Figref{fig:core_comparison} shows the evolution of the nuclear and
neutrino luminosities (integrated throughout the entire star) and the
central temperature of these two models. Even if there are some
differences, the rough timing and amplitude of the oscillations of the
core quantities are not significantly different during the pulses,
i.e.~during the last year of evolution. This is expected, since in
general for all evolved massive stars the core evolution is not
determined by the envelope. Instead, the evolution is driven by the
neutrino losses from the core itself, rather than the photon
luminosity at the surface \citep[e.g.,][]{fraley:68}.

\begin{figure}[htbp!]
  \centering
  \includegraphics[width=0.5\textwidth]{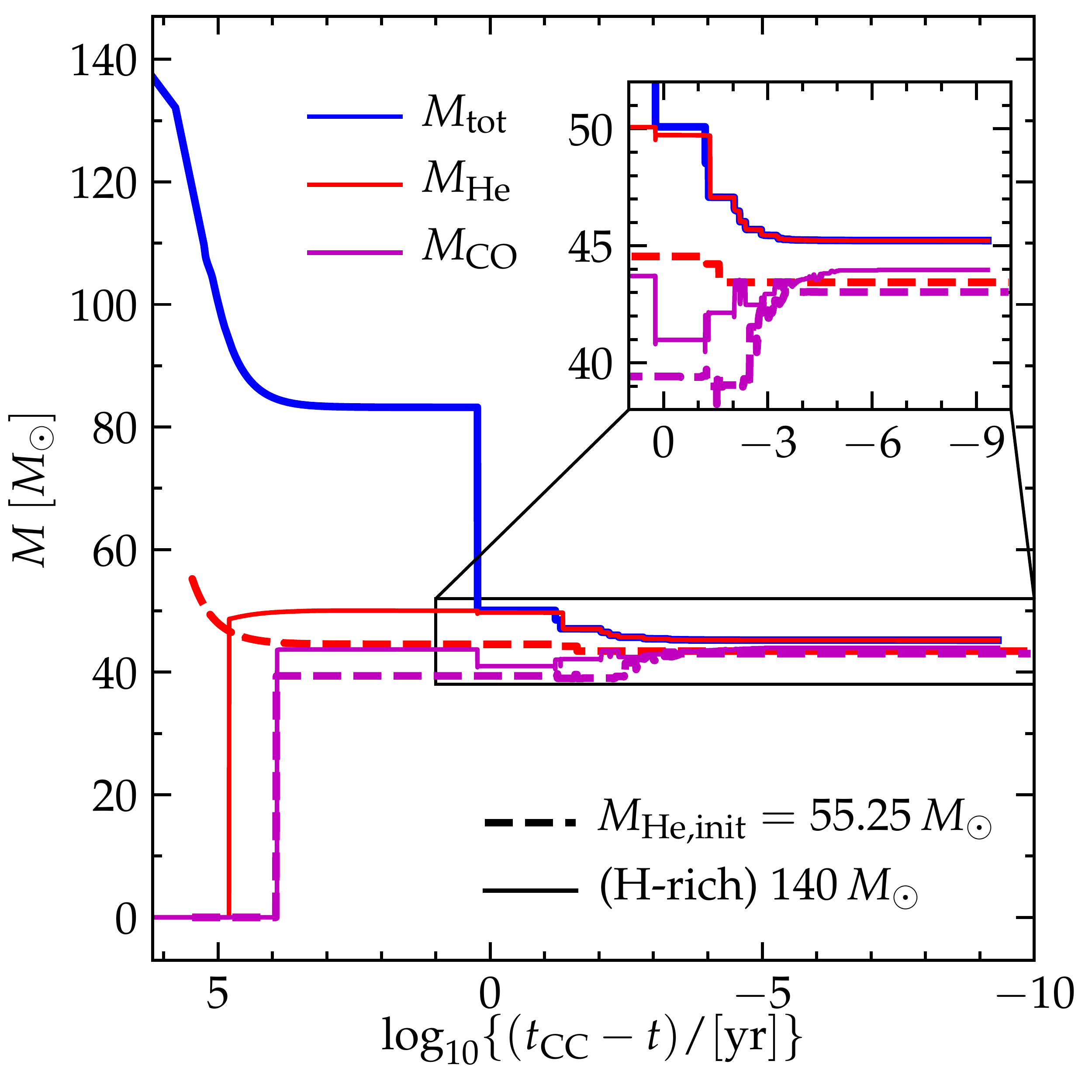}
  \caption{Comparison of the evolution of the total mass (blue), He
    core mass (red) and CO core mass (purple) for a $140\,M_\odot$
    H-rich model (solid) and a $M_\mathrm{He, init}=55.25\,M_\odot$
    He-core (dashed).}
  \label{fig:mass_comparison}
\end{figure}

\Figref{fig:mass_comparison} shows the evolution of the total, He and
CO core masses for these two
models. For the H-rich model, we define the He core as the outermost
location where the H mass fraction is below 0.5 and the He mass
fraction exceeds 0.1. If this yields zero, we set the He core mass
equal to the total core mass. To avoid noise in the curve caused by
mixing episodes at the core boundary at the end of the main sequence,
we only show the He core mass once its value has stabilized. The H-rich model
expells all its envelope at the first pulse about one year before
CC. At this point the He core and the total mass become the same.

As shown in Table~\ref{tab:ejecta}, our $55.25\,M_\odot$ model has
three distinct mass-loss events. The comparison H-rich model also
exhibits three mass-loss pulses. However, the timing in between the events is
slightly different in the two models. This is partly due to the extreme
sensitivity to the core masses (or, more precisely, the entropy
profile) at the onset of the pulses. As well as the presence
of the H-rich envelope, especially if extended, which increases the
dynamical timescale and the delay between the core thermonuclear
ignition and shock breakout and thus the mass ejection.

These two models, characterized by similar $M_\mathrm{CO}$ (defined in
\Secref{sec:methods} as the maximum throughout the evolution of the CO
core mass) yield similar BH masses of $\simeq43.5\,M_\odot$ for the He
core and $\simeq45\,M_\odot$ for the H-rich model. The difference
between the BH mass values found here are smaller than those
introduced by other physical and numerical uncertainties
\citep[][]{farmer:19}.

\section{Mass loss per pulse}

Table~\ref{tab:ejecta} provides data for each individual mass ejection event in
our grid of models. We list the total mass at the beginning of the
pulse $M_\mathrm{tot}^\mathrm{pre-pulse}$, the amount of mass lost 
$\Delta M_\mathrm{pulse}$, the delay time between core-collapse and
the end of the pulse, and the pulse duration $\Delta t_\mathrm{pulse}$,
 respectively (as defined in \Secref{sec:mejection_def}), and the center-of-mass velocity of the ejected layer
 $\langle v \rangle$. This table is also available at \href{https://doi.org/10.5281/zenodo.3406356}{https://doi.org/10.5281/zenodo.3406356}.

\tablehead{%
  \hline\hline
  $M_\mathrm{He,init}$ & pulse & $M_\mathrm{tot}^\mathrm{pre-pulse}$ &
  $\Delta M_\mathrm{pulse}$ & $\log_{10}(t_\mathrm{CC}-t_\mathrm{pulse\ end})$ & $\Delta t_\mathrm{pulse}$ &
  $\langle v \rangle$ \\
  $[M_\odot]$ & & $[M_\odot]$ & $[M_\odot]$ & [yr] & [hours] & $\mathrm{[10^3 km \ s^{1}]}$ \\\hline\hline
}

\tabletail{%
  \hline
  \multicolumn{7}{c}{{Continued on next column}}\\
  \hline
}

\tablelasttail{%
  \hline
  \multicolumn{7}{c}{{Concluded}} \\
\hline}

\bottomcaption{Number of pulses (pulse), prepulse total mass
$M_\mathrm{tot}^\mathrm{pre-pulse}$, amount of mass lost in the pulse
$\Delta M_\mathrm{pulse}$, delay time between the pulse end $t_\mathrm{end}$ and the
final core-collapse and
its duration $\Delta t_\mathrm{pulse}$ (so that 
$t_\mathrm{pulse\ end}=t_\mathrm{pulse\ start}+\Delta t_\mathrm{pulse}$, cf. \Figref{fig:timing}) according to the definition of
\Secref{sec:mejection_def}, and the velocity of the center of mass of the
ejected layers $\langle v\rangle$ for each PPI+CC model. \label{tab:ejecta}
}

\begin{center}
  \begin{tiny}
    \setlength{\tabcolsep}{1pt}
    \begin{supertabular}{c|c|c|c|c|c|c}
      \centering
      \input{pulses_table_new.txt}
    \end{supertabular}
  \end{tiny}  
\end{center}

\end{document}

%% file: pulses_table_new.txt
41.25  &   1  &   34.60  &   0.00  &   -6.14  &   0  &   3.83\\
\hline42.00  &   1  &   35.14  &   0.00  &   -4.79  &   0  &   2.84\\
\hline42.25  &   1  &   35.33  &   0.00  &   -4.21  &   1  &   2.45\\
\hline42.50  &   1  &   35.51  &   0.01  &   -3.84  &   1  &   1.97\\
\hline42.75  &   1  &   35.69  &   0.02  &   -3.75  &   2  &   1.87\\
\hline43.00  &   1  &   35.87  &   0.01  &   -3.74  &   2  &   1.85\\
\hline43.25  &   1  &   36.05  &   0.02  &   -3.59  &   2  &   1.77\\
\hline43.50  &   1  &   36.23  &   1.52  &   -1.92  &   105  &   3.87\\
\hline43.75  &   1  &   36.42  &   0.55  &   -2.32  &   33  &   3.52\\
\hline44.00  &   1  &   36.59  &   0.87  &   -2.06  &   73  &   3.63\\
\hline44.25  &   1  &   36.77  &   0.54  &   -2.21  &   53  &   3.27\\
\hline44.50  &   1  &   36.95  &   0.05  &   -3.37  &   4  &   1.72\\
\hline44.75  &   1  &   37.13  &   0.06  &   -3.15  &   6  &   1.64\\
\hline45.00  &   1  &   37.31  &   0.07  &   -3.08  &   7  &   1.64\\
\hline45.25  &   1  &   37.50  &   0.07  &   -3.01  &   8  &   1.61\\
\hline45.50  &   1  &   37.67  &   0.09  &   -3.01  &   9  &   1.57\\
\hline45.75  &   1  &   37.85  &   0.05  &   -2.89  &   11  &   1.46\\
\hline46.00  &   1  &   38.03  &   0.06  &   -2.85  &   12  &   1.55\\
\hline46.25  &   1  &   38.21  &   0.03  &   -2.81  &   11  &   1.37\\
\hline46.50  &   1  &   38.39  &   0.03  &   -2.80  &   10  &   1.40\\
\hline46.75  &   1  &   38.56  &   0.13  &   -2.74  &   16  &   1.59\\
\hline47.00  &   1  &   38.74  &   0.09  &   -2.72  &   17  &   1.54\\
\hline47.25  &   1  &   38.92  &   0.29  &   -2.62  &   21  &   1.66\\
\hline47.50  &   1  &   39.10  &   0.44  &   -2.46  &   30  &   1.64\\
\hline47.75  &   1  &   39.28  &   0.03  &   -2.73  &   15  &   1.36\\
\hline48.00  &   1  &   39.45  &   0.61  &   -2.23  &   51  &   1.78\\
\hline48.25  &   1  &   39.64  &   0.48  &   -2.36  &   38  &   1.56\\
\hline48.75  &   1  &   39.98  &   0.89  &   -2.06  &   76  &   1.61\\
\hline49.00  &   1  &   40.17  &   1.08  &   -1.76  &   152  &   1.63\\
\hline49.25  &   1  &   40.34  &   1.14  &   -1.74  &   160  &   1.62\\
\hline49.50  &   1  &   40.52  &   0.52  &   -2.30  &   43  &   1.78\\
\hline49.75  &   1  &   40.70  &   1.64  &   -1.69  &   180  &   1.66\\
\hline50.00  &   1  &   40.88  &   1.24  &   -1.84  &   127  &   1.67\\
\hline50.25  &   1  &   41.05  &   1.84  &   -1.63  &   208  &   1.70\\
\hline50.50  &   1  &   41.23  &   1.69  &   -1.78  &   145  &   1.78\\
\hline50.75  &   1  &   41.41  &   1.97  &   -1.62  &   213  &   1.75\\
\hline\multirow{2}{*}{51.00}  &   1  &   41.58  &   0.40  &   -1.27  &   30  &   2.21\\
  &   2  &   41.18  &   1.61  &   -2.74  &   16  &   1.45\\
\hline
51.50  &   1  &   41.93  &   1.85  &   -1.58  &   231  &   1.88\\
\hline\multirow{2}{*}{51.75}  &   1  &   42.11  &   0.43  &   -1.25  &   35  &   2.11\\
  &   2  &   41.68  &   1.70  &   -2.39  &   36  &   1.64\\
\hline
\multirow{2}{*}{52.00}  &   1  &   42.28  &   0.74  &   -1.10  &   65  &   2.22\\
  &   2  &   41.53  &   1.49  &   -2.78  &   15  &   1.51\\
\hline
52.25  &   1  &   42.46  &   0.15  &   -1.90  &   110  &   1.92\\
\hline52.50  &   1  &   42.64  &   0.09  &   -1.83  &   131  &   1.60\\
\hline52.75  &   1  &   42.81  &   0.39  &   -1.71  &   169  &   1.86\\
\hline53.00  &   1  &   42.98  &   0.24  &   -1.86  &   120  &   1.51\\
\hline53.25  &   1  &   43.16  &   0.12  &   -2.13  &   65  &   1.59\\
\hline53.50  &   1  &   43.33  &   1.41  &   -1.50  &   280  &   1.99\\
\hline53.75  &   1  &   43.50  &   1.23  &   -1.56  &   240  &   1.95\\
\hline54.00  &   1  &   43.68  &   1.56  &   -1.32  &   417  &   1.81\\
\hline54.50  &   1  &   44.03  &   1.75  &   -1.43  &   328  &   1.95\\
\hline\multirow{2}{*}{54.75}  &   1  &   44.20  &   1.29  &   -1.23  &   203  &   2.12\\
  &   2  &   42.90  &   0.56  &   -2.57  &   24  &   1.51\\
\hline
55.00  &   1  &   44.37  &   1.19  &   -1.37  &   372  &   1.97\\
\hline\multirow{3}{*}{55.25}  &   1  &   44.55  &   0.32  &   -1.24  &   3  &   2.19\\
  &   2  &   44.23  &   0.78  &   -1.24  &   282  &   1.85\\
  &   3  &   43.45  &   0.25  &   -2.26  &   48  &   1.49\\
\hline
55.50  &   1  &   44.72  &   1.32  &   -1.40  &   351  &   1.88\\
\hline\multirow{2}{*}{55.75}  &   1  &   44.89  &   0.44  &   -1.21  &   4  &   2.14\\
  &   2  &   44.45  &   1.06  &   -1.76  &   152  &   1.68\\
\hline
\multirow{2}{*}{56.00}  &   1  &   45.07  &   0.50  &   -1.06  &   3  &   2.16\\
  &   2  &   44.57  &   1.10  &   -1.58  &   231  &   1.76\\
\hline
\multirow{2}{*}{56.25}  &   1  &   45.24  &   0.52  &   -1.07  &   3  &   2.18\\
  &   2  &   44.72  &   1.13  &   -1.74  &   161  &   1.59\\
\hline
\multirow{2}{*}{56.50}  &   1  &   45.42  &   0.64  &   -0.94  &   6  &   2.14\\
  &   2  &   44.77  &   0.75  &   -1.98  &   91  &   1.48\\
\hline
\multirow{2}{*}{56.75}  &   1  &   45.59  &   0.63  &   -0.95  &   6  &   2.14\\
  &   2  &   44.96  &   1.81  &   -1.81  &   135  &   1.79\\
\hline
\multirow{2}{*}{57.00}  &   1  &   45.76  &   0.69  &   -0.84  &   5  &   2.16\\
  &   2  &   45.07  &   1.39  &   -0.85  &   1249  &   1.66\\
\hline
\multirow{2}{*}{57.25}  &   1  &   45.93  &   0.74  &   -0.78  &   9  &   2.14\\
  &   2  &   45.19  &   1.58  &   -0.78  &   1449  &   1.67\\
\hline
\multirow{2}{*}{57.50}  &   1  &   46.10  &   0.89  &   -0.57  &   17  &   2.14\\
  &   2  &   45.21  &   1.31  &   -1.93  &   103  &   1.54\\
\hline
\multirow{2}{*}{57.75}  &   1  &   46.27  &   0.99  &   -0.41  &   30  &   2.15\\
  &   2  &   45.28  &   2.22  &   -1.60  &   218  &   1.96\\
\hline
\multirow{2}{*}{58.00}  &   1  &   46.45  &   1.09  &   -0.26  &   29  &   2.16\\
  &   2  &   45.35  &   2.29  &   -1.57  &   234  &   1.65\\
\hline
\multirow{2}{*}{58.25}  &   1  &   46.62  &   1.16  &   -0.15  &   34  &   2.16\\
  &   2  &   45.45  &   2.90  &   -0.15  &   6239  &   1.77\\
\hline
\multirow{2}{*}{58.50}  &   1  &   46.79  &   1.08  &   -0.30  &   34  &   2.15\\
  &   2  &   45.71  &   1.70  &   -1.77  &   150  &   1.63\\
\hline
\multirow{2}{*}{58.75}  &   1  &   46.96  &   1.29  &   0.03  &   38  &   2.18\\
  &   2  &   45.67  &   1.68  &   -1.25  &   496  &   1.76\\
\hline
\multirow{2}{*}{59.00}  &   1  &   47.13  &   1.18  &   -0.14  &   34  &   2.16\\
  &   2  &   45.95  &   3.26  &   -0.14  &   6323  &   1.85\\
\hline
\multirow{2}{*}{59.25}  &   1  &   47.31  &   1.45  &   0.27  &   38  &   2.20\\
  &   2  &   45.86  &   1.69  &   -1.15  &   451  &   1.87\\
\hline
\multirow{3}{*}{59.50}  &   1  &   47.48  &   1.31  &   0.04  &   36  &   2.18\\
  &   2  &   46.17  &   1.04  &   -1.34  &   147  &   1.71\\
  &   3  &   44.19  &   0.47  &   -3.10  &   7  &   1.70\\
\hline
\multirow{2}{*}{59.75}  &   1  &   47.65  &   1.50  &   0.32  &   38  &   2.20\\
  &   2  &   46.15  &   1.61  &   -1.09  &   242  &   1.63\\
\hline
\multirow{3}{*}{60.25}  &   1  &   48.00  &   1.38  &   0.14  &   38  &   2.18\\
  &   2  &   46.61  &   1.80  &   -1.26  &   228  &   1.66\\
  &   3  &   44.24  &   0.02  &   -4.15  &   1  &   1.73\\
\hline
\multirow{2}{*}{60.75}  &   1  &   48.33  &   1.61  &   0.47  &   39  &   2.20\\
  &   2  &   46.71  &   2.06  &   -1.28  &   458  &   1.58\\
\hline
\multirow{2}{*}{61.75}  &   1  &   49.00  &   1.80  &   0.71  &   44  &   2.23\\
  &   2  &   47.20  &   1.26  &   -1.55  &   247  &   1.69\\
\hline
\multirow{2}{*}{62.75}  &   1  &   49.67  &   2.14  &   1.21  &   49  &   2.29\\
  &   2  &   47.53  &   1.91  &   -1.46  &   305  &   1.79\\
\hline
\multirow{2}{*}{63.75}  &   1  &   50.34  &   2.66  &   1.98  &   84  &   2.39\\
  &   2  &   47.67  &   1.56  &   -1.78  &   147  &   1.51\\
\hline
\multirow{2}{*}{64.25}  &   1  &   50.67  &   2.72  &   2.04  &   78  &   2.39\\
  &   2  &   47.94  &   1.46  &   -1.68  &   185  &   1.50\\
\hline
\multirow{2}{*}{64.75}  &   1  &   51.01  &   2.99  &   2.43  &   100  &   2.47\\
  &   2  &   48.00  &   1.25  &   -1.74  &   160  &   1.34\\
\hline
\multirow{3}{*}{65.75}  &   1  &   51.66  &   3.44  &   2.89  &   187  &   2.58\\
  &   2  &   48.16  &   0.01  &   -1.57  &   43  &   0.72\\
  &   3  &   44.89  &   1.11  &   -2.43  &   32  &   1.98\\
\hline
\multirow{2}{*}{66.25}  &   1  &   52.00  &   3.64  &   3.03  &   332  &   2.67\\
  &   2  &   48.27  &   1.16  &   -1.53  &   257  &   1.17\\
\hline
\multirow{2}{*}{66.75}  &   1  &   52.32  &   3.75  &   3.14  &   165  &   2.81\\
  &   2  &   48.47  &   2.55  &   -1.52  &   267  &   1.11\\
\hline
\multirow{3}{*}{67.25}  &   1  &   52.65  &   3.80  &   3.22  &   113  &   2.96\\
  &   2  &   48.71  &   0.00  &   -1.17  &   33  &   0.56\\
  &   3  &   44.40  &   1.22  &   -2.08  &   73  &   1.92\\
\hline
\multirow{2}{*}{67.75}  &   1  &   52.99  &   3.91  &   3.22  &   128  &   2.89\\
  &   2  &   45.20  &   1.57  &   -1.60  &   219  &   2.11\\
\hline
\multirow{3}{*}{68.25}  &   1  &   53.31  &   3.97  &   3.35  &   254  &   3.25\\
  &   2  &   49.15  &   0.00  &   -1.11  &   6  &   0.47\\
  &   3  &   44.24  &   1.35  &   -2.31  &   43  &   1.78\\
\hline
\multirow{2}{*}{68.75}  &   1  &   53.63  &   3.94  &   3.36  &   221  &   3.33\\
  &   2  &   44.54  &   1.52  &   -1.99  &   88  &   1.86\\
\hline
\multirow{2}{*}{69.25}  &   1  &   53.97  &   4.08  &   3.43  &   226  &   3.69\\
  &   2  &   44.08  &   1.64  &   -1.95  &   95  &   1.92\\
\hline
\multirow{3}{*}{69.75}  &   1  &   54.29  &   4.20  &   3.47  &   390  &   3.86\\
  &   2  &   44.31  &   0.47  &   -1.25  &   271  &   2.34\\
  &   3  &   44.31  &   1.84  &   -1.25  &   488  &   2.15\\
\hline
\multirow{2}{*}{70.25}  &   1  &   54.62  &   4.72  &   3.51  &   176  &   3.90\\
  &   2  &   43.59  &   1.62  &   -1.88  &   95  &   1.87\\
\hline
\multirow{3}{*}{70.75}  &   1  &   54.94  &   4.89  &   3.53  &   218  &   3.89\\
  &   2  &   49.77  &   0.00  &   -0.88  &   49  &   0.51\\
  &   3  &   43.49  &   0.80  &   -1.98  &   87  &   2.11\\
\hline
\multirow{3}{*}{71.00}  &   1  &   55.11  &   5.02  &   3.54  &   196  &   3.89\\
  &   2  &   49.80  &   0.00  &   -0.80  &   83  &   0.58\\
  &   3  &   43.41  &   0.82  &   -2.07  &   74  &   2.13\\
\hline
\multirow{3}{*}{71.25}  &   1  &   55.26  &   5.73  &   3.58  &   125  &   3.86\\
  &   2  &   49.22  &   0.01  &   -0.85  &   130  &   0.63\\
  &   3  &   42.61  &   0.79  &   -2.10  &   69  &   2.15\\
\hline
\multirow{3}{*}{71.50}  &   1  &   55.42  &   5.39  &   3.56  &   78  &   3.88\\
  &   2  &   49.73  &   0.01  &   -0.78  &   134  &   0.68\\
  &   3  &   43.06  &   0.72  &   -2.45  &   31  &   2.02\\
\hline
\multirow{3}{*}{71.75}  &   1  &   55.59  &   5.97  &   3.59  &   294  &   3.84\\
  &   2  &   49.31  &   0.01  &   -0.83  &   161  &   0.72\\
  &   3  &   42.48  &   0.48  &   -3.05  &   8  &   1.77\\
\hline
\multirow{3}{*}{72.00}  &   1  &   55.74  &   5.85  &   3.59  &   236  &   3.85\\
  &   2  &   49.58  &   0.01  &   -0.78  &   173  &   0.77\\
  &   3  &   42.63  &   0.73  &   -2.28  &   46  &   2.25\\
\hline
\multirow{3}{*}{72.25}  &   1  &   55.91  &   7.37  &   3.63  &   603  &   3.63\\
  &   2  &   48.19  &   0.01  &   -0.92  &   194  &   0.81\\
  &   3  &   41.57  &   0.42  &   -3.12  &   7  &   1.83\\
\hline
\multirow{3}{*}{72.50}  &   1  &   56.07  &   7.32  &   3.63  &   656  &   3.63\\
  &   2  &   48.41  &   1.30  &   -0.88  &   243  &   0.71\\
  &   3  &   41.66  &   0.54  &   -2.94  &   10  &   1.94\\
\hline
\multirow{3}{*}{72.75}  &   1  &   56.24  &   7.77  &   3.64  &   891  &   3.59\\
  &   2  &   48.11  &   1.35  &   -0.88  &   273  &   0.73\\
  &   3  &   41.39  &   0.59  &   -2.77  &   15  &   2.05\\
\hline
\multirow{3}{*}{73.00}  &   1  &   56.39  &   8.29  &   3.66  &   1444  &   3.55\\
  &   2  &   42.48  &   0.38  &   -0.93  &   281  &   2.55\\
  &   3  &   42.48  &   1.74  &   -0.93  &   1030  &   2.26\\
\hline
\multirow{3}{*}{73.25}  &   1  &   56.55  &   8.91  &   3.67  &   1746  &   3.48\\
  &   2  &   42.16  &   0.44  &   -0.97  &   316  &   2.52\\
  &   3  &   42.16  &   1.75  &   -0.97  &   938  &   2.27\\
\hline
\multirow{3}{*}{73.50}  &   1  &   56.72  &   9.98  &   3.70  &   2120  &   3.43\\
  &   2  &   46.34  &   1.31  &   -1.02  &   265  &   0.75\\
  &   3  &   40.11  &   0.47  &   -2.76  &   15  &   2.02\\
\hline
\multirow{3}{*}{73.75}  &   1  &   56.87  &   10.38  &   3.70  &   3240  &   3.43\\
  &   2  &   46.09  &   1.30  &   -1.04  &   266  &   0.75\\
  &   3  &   39.92  &   0.45  &   -2.80  &   14  &   2.01\\
\hline
\multirow{3}{*}{74.75}  &   1  &   57.51  &   12.68  &   3.74  &   3825  &   3.37\\
  &   2  &   44.46  &   0.01  &   -1.23  &   215  &   0.83\\
  &   3  &   38.68  &   0.43  &   -2.63  &   21  &   2.08\\
\hline
\multirow{3}{*}{75.00}  &   1  &   57.68  &   13.47  &   3.75  &   1685  &   3.34\\
  &   2  &   44.20  &   5.50  &   3.75  &   49697937  &   0.80\\
  &   3  &   38.34  &   1.27  &   -1.67  &   187  &   2.33\\
\hline
\multirow{3}{*}{75.25}  &   1  &   57.84  &   13.33  &   3.75  &   1850  &   3.35\\
  &   2  &   44.51  &   5.66  &   3.75  &   49472412  &   0.86\\
  &   3  &   38.49  &   1.67  &   -1.74  &   160  &   2.43\\
\hline
\multirow{3}{*}{75.50}  &   1  &   57.99  &   13.89  &   3.76  &   1400  &   3.31\\
  &   2  &   44.10  &   5.49  &   3.76  &   50205810  &   0.85\\
  &   3  &   38.25  &   1.30  &   -1.66  &   192  &   2.35\\
\hline
\multirow{2}{*}{76.25}  &   1  &   58.47  &   15.93  &   3.78  &   1875  &   3.28\\
  &   2  &   42.18  &   1.72  &   -1.39  &   358  &   1.03\\
\hline
\multirow{2}{*}{76.50}  &   1  &   58.63  &   16.99  &   3.79  &   542  &   3.23\\
  &   2  &   41.28  &   1.69  &   -1.48  &   289  &   0.99\\
\hline
\multirow{2}{*}{76.75}  &   1  &   58.79  &   18.06  &   3.80  &   888  &   3.19\\
  &   2  &   40.39  &   1.61  &   -1.64  &   201  &   0.92\\
\hline
\multirow{2}{*}{77.00}  &   1  &   58.96  &   19.04  &   3.81  &   1106  &   3.15\\
  &   2  &   39.57  &   2.26  &   -1.79  &   144  &   1.02\\
\hline
\multirow{2}{*}{77.25}  &   1  &   59.10  &   20.35  &   3.82  &   958  &   3.09\\
  &   2  &   38.43  &   1.99  &   -2.00  &   88  &   1.14\\
\hline
\multirow{2}{*}{77.50}  &   1  &   59.27  &   21.33  &   3.82  &   991  &   3.05\\
  &   2  &   37.61  &   1.75  &   -2.09  &   72  &   1.03\\
\hline
\multirow{2}{*}{77.75}  &   1  &   59.43  &   22.45  &   3.83  &   1234  &   3.01\\
  &   2  &   36.66  &   1.55  &   -2.25  &   49  &   0.99\\
\hline
\multirow{2}{*}{80.00}  &   1  &   60.84  &   41.62  &   3.99  &   20748  &   2.35\\
  &   2  &   19.03  &   0.01  &   -4.41  &   0  &   2.55\\
\hline